\documentclass[alpha-refs]{wiley-article}

\usepackage{ulem}
\usepackage{amsmath}%
\usepackage{amsfonts}%
\usepackage{amssymb}%
\usepackage{graphicx}
\usepackage{amssymb}
\usepackage{color}
\usepackage{tikz} 
\usepackage[autostyle=true,german=quotes]{csquotes} 
\usepackage{framed} 
\usepackage{subcaption} 
\usepackage{nicefrac} 
\usepackage{bbm} 
\usepackage{float} 
\usepackage{booktabs} 
\usepackage{multirow} 
\usepackage{multicol} 
\usepackage{nicefrac}

\usepackage{mathtools} 
\mathtoolsset{showonlyrefs}

\newcommand{\bs}{\boldsymbol}
\newcommand{\E}{\mathbb{E}}
\newcommand{\Var}{\mathbb{V}\text{ar}}
\newcommand{\Cov}{\mathbb{C}\text{ov}}

\newcommand{\prob}{\mathbb{P}}
\newcommand{\nullhyp}{\text{H}_0:}
\newcommand{\alterhyp}{\text{H}_1:}
\newcommand{\Gnp}{\mathcal{G}_{ER}(n)}
\newcommand{\GnP}{\mathcal{G}_{HER}(n)}
\newcommand{\ntoinfty}{\xrightarrow[n\to\infty]{}}
\newcommand{\convD}{\xrightarrow{~d~}}
\newcommand{\convP}{\xrightarrow{~P~}}
\newcommand{\Pmat}{\bs{P}}
\newcommand{\landau}{\mathcal{O}}
\newcommand{\ph}{\hat{p}}

\newcommand{\pra}{p_\text{intra}}
\newcommand{\per}{p_\text{inter}}
\newcommand{\bin}{\binom{n}{3}}
\newcommand{\binHalf}{\binom{\nicefrac{n}{2}}{3}}
\newcommand{\binHalb}{\binom{\nicefrac{n}{2}}{2}}

\usepackage{siunitx}

\papertype{Original Article}
\paperfield{Journal Section}

\title{Goodness-of-fit testing based on graph functionals for homogeneous Erdös-Rényi graphs}

\author[1]{Barbara Brune}
\author[2]{Jonathan Flossdorf}
\author[2]{Carsten Jentsch}

\affil[1]{Institute of Statistics and Mathematical Methods in Economics, TU Wien, Vienna, Austria}
\affil[2]{Department of Statistics, TU Dortmund University, Dortmund, Germany}

\corraddress{Jonathan Flossdorf, Department of Statistics, TU Dortmund University, Vogelpothsweg 87, 44221 Dortmund, Germany}
\corremail{flossdorf@statistik.tu-dortmund.de}

\fundinginfo{Mercator Research Center Ruhr (MERCUR), Grant Number: PR-2019-0019}

\begin{document}

\maketitle

\begin{abstract}
The Erdös‐Rényi graph is a popular choice to model network data as it is parsimoniously parametrized, straightforward to interpret and easy to estimate. However, it has limited suitability in practice, since it often fails to capture crucial characteristics of real-world networks. To check its adequacy, we propose a novel class of goodness-of-fit tests for homogeneous Erdös‐Rényi models against heterogeneous alternatives that permit non-constant edge probabilities. In this context, we allow for both asymptotically dense and sparse networks. The tests are based on graph functionals that cover a broad class of network statistics for which we derive limiting distributions in a unified manner. The resulting class of asymptotic tests includes several existing tests as special cases. Further, we propose a parametric bootstrap and prove its consistency, which enables performance improvements particularly for small network sizes and avoids the often tedious variance estimation for asymptotic tests. Moreover, we analyse the sensitivity of different goodness-of-fit test statistics that rely on popular choices of subgraphs. We evaluate the proposed class of tests and illustrate our theoretical findings by extensive simulations.

\end{abstract}

\section{Introduction}
Due to the technical progress in recent decades, not only the amount of data is growing but also its complexity. Hence, the analysis of statistical network data has become a popular research objective with applications ranging from social sciences \citep{sarkar2005dynamic, carrington2005models} or biology \citep{bassett2017network, prill2005dynamic} to logistics and transportation processes \citep{lee2009dynamic}. Various mathematical and computational methods have been developed to analyze, model, and understand the behaviour of networks. As network data is rather complex by nature, it is a typical first step to find suitable complexity reduction methods. A common approach is to calculate various sorts of summary statistics from the observed network such as matrix norms or centrality measures that may describe the network structure from different perspectives. However, the general difficulty is to be aware of the limited information that those statistics are able to capture \citep{flossdorf2021change,flossdorf2022online}, since they reduce the information of a whole network to a few scalar values. Another approach is to fit a suitable network model to the observed data in order to analyze graphs in a simplified and controlled setup and to perform statistical inference. One of the simplest network models is the \textsc{Erdös-Rényi model} \citep{gilbert1959random} in which the edges are considered as independent Bernoulli variables with a common probability parameter $p$. Obviously, this model may have substantial shortcomings in modeling real world networks. Due to the assumption that all edges form independently with the same probability, it usually fails to capture many of their features, e.g.~in social networks two people are more likely to know each other if they have a common friend. This yields clusters of vertices that have a higher edge probability between each other than to other vertices. A solution for this is to fit more general random graph models that allow for varying edge probabilities leading to the class of heterogenous Erdös-Rényi models \citep{ouadah2020degree} with the stochastic block model (SBM) \citep{holland1983stochastic} as a popular special case. Whereas these heterogenous models might match the features of real world networks noticeably better than their homogenous counterparts, their rigorous analysis is much more cumbersome from a graph theory point of view. From a practitioner's perspective, it is desirable to choose the most parsimonious model with as few parameters as possible in order to simplify the estimation process which is why the homogenous Erdös-Rényi model with only one probability parameter is still a popular benchmark model. Therefore, it is a crucial question if the usage of a parsimonious model is justified or if a more complex model achieves a significantly better fit. This leads to the field of goodness-of-fit procedures that aims to decide whether a specified model fits the underlying data adaquately or if another alternative would be a better choice.    

Several approaches for goodness-of-fit testing for random graphs have been proposed in the literature. This involves tests for the number of communities in stochastic block models which are based on the largest singular value of a residual matrix that is obtained by the difference of the estimated block mean effect and the adjacency matrix \citep{lei2016goodness}. Similarly, the principal eigenvalue \citep{bickel2016hypothesis} or the maximum entry \citep{hu2021using} of a centered and scaled adjacency matrix is used to test whether a random graph is generated by a homogenous Erdös-Rényi model or a stochastic block model. For the same hypotheses, \cite{gao2017testing} developed an asymptotic test based on specific subgraph counts. Subgraph counts as a test statistic are also used in \cite{maugis2017statistical} under the rather restrictive and often unrealistic assumption that multiple independent and identically distributed samples of graphs of the same size are available, and in \cite{ospina2019assessment} within a Monte-Carlo framework. A more general goodness-of-fit method is proposed in \cite{dan2020goodness}, where the goal is to test whether the probability matrix of independent samples of a heterogenous graph matches a specified reference matrix that has to be assumed to be known. To do so, optimal minimax sample complexities in various matrix norms such as e.g.~the Frobenius norm are derived. In \cite{ouadah2020degree}, goodness-of-fit tests for Erdös-Rényi-type models have been derived based on the degree variance statistic that serves as an heterogeneity index of a graph \citep{snijders1981degree}. A statistical test for exponential random graph models \citep{robins2007introduction} is proposed in \cite{xu2021stein} with test statistics that are derived from a kernel Stein discrepancy. Furthermore, a goodness-of-fit method for stochastic actor oriented models \citep{snijders1996stochastic} is presented in \cite{lospinoso2019goodness}.

We investigate a unified approach for goodness-of-fit testing for homogeneous Erdös-Rényi models. Precisely, we propose tests that aim to tell whether an observed network was either generated by some homogenous Erdös-Rényi model or by some other, more heterogenous model. Contrary to the related literature, we do not limit our analysis to a particular test statistic, but use the concept of graph functionals which allows for a unified treatment in theory and practice. Precisely, we propose a class of test statistics that covers a wide range of popular network metrics such as the degree variance statistic, average centrality metrics, as well as subgraph counts of arbitrary order and shape. Hence, our approach is very flexible  
and contains various already proposed tests \citep{gao2017testing, ouadah2020degree} as special cases. Following this approach, we derive general asymptotic theory for the whole proposed class of test statistics in a unified manner. This particularly avoids the tedious derivation of the individual asymptotics for newly proposed test statistics that fall into our framework in a case by case manner. Furthermore, while the implementation of asymptotic tests is often tedious as their variances are highly case-dependent, we also propose a parametric bootstrap approach and prove a general bootstrap consistency result in order to develop bootstrap versions of the goodness-of-fit tests. These enable us to gain finite sample improvements and to circumvent the tedious variance estimation.

The paper is organized as follows. It starts off with the necessary graph theory concepts in Section \ref{sec:theory}, including theory on graph functionals and random graphs. Particularly, we recap important traits of graph functionals and extend them in order to enable the derivation of general asymptotic theory and the development of asymptotic goodness-of-fit tests for these class of statistics. In Section \ref{sec:boots}, we derive the required theory for enabling the construction of parametric bootstrap versions of this test. Based on this theory, we construct the novel class of goodness-of-fit tests both in an asymptotic and a bootstrap manner in Section \ref{sec:testing}. We give detailed examples and investigate the suitability of several specific graph functionals for goodness-of-fit testing in the case of SBM alternatives. Subsequently, we evaluate the performance of the proposed procedure in an extensive simulation study in Section \ref{sec:Sim}. The final Section \ref{sec:concl} consists of some concluding remarks. Most proofs and additional results are deferred to the Appendix.

\section{Random Graphs and Graph Functionals}\label{sec:theory}
In this chapter, we give an overview of the used network generating models and the concept of graph functionals. Particularly, we gather and extend existing limiting distribution theory for this class of statistics in order to enable the development of unified asymptotic and bootstrap goodness-of-fit tests based on graph functionals.

\subsection{Settings}
Suppose we observe a random graph $G$ that is defined by a \textsc{vertex set} $V=V(G)$, and \textsc{edge set} $E=E(G)$. The edge set consists of pairs of vertices $\{v_1,v_2\}$, where $v_1,v_2\in V(G)$. We denote by $n:=n(G)=|V(G)|$ the number of vertices, and by $m:=m(G)=|E(G)|$ the number of edges of the graph $G$, respectively. The vertex set is denoted by $V(G)=\{v_1,...,v_n\}$, and each edge in $E(G)$ connects two of the vertices. The network can alternatively be represented by an adjacency matrix $A=(A_{ij})$ of dimension $(n \times n)$. In this paper, we concentrate on unweighted and undirected networks without self-loops, which are often called simple graphs. Their adjacency matrices are binary and symmetric such that $A_{ij}=1$ indicates the presence of an edge between two vertices $i$ and $j$, and $A_{ij} = 0$ if there is no edge between the respective vertices. As self-loops are not allowed, we have $A_{ii} = 0$ for all $i$. Throughout the paper, we focus on goodness-of-fit testing for random graphs. As mentioned, we use the well-established Erdös-Rényi graph as the benchmark model.
\begin{definition}[Definition 1]\textbf{[{Erdös-Rényi \textbf{(ER)} graph}]} \label{def:ER}
Let $G=(V,E)$ be a random graph on $n$ vertices with adjacency matrix $A$ and let $p\in[0,1]$ be the connection probability. Then, we call $G$ an \textit{Erdös-Rényi graph} $\mathcal{G}(n,p)$, if the edges are realizations of stochastically independent and identically Bernoulli distributed random variables. That is, for $1\leq i < j \leq n$, we have $A_{ij}\sim \mathnormal{\textup{Bin}}(1,p)$ with $A_{ji}:=A_{ij}$, and $A_{ii}:=0$ for all $i$. We denote the resulting ER model class by
\begin{align*}
\mathcal{G}_{ER}(n)=\{\mathcal{G}(n,p),p\in[0,1]\}.
\end{align*}
\end{definition} 

As the ER-model has quite restrictive assumptions, various generalizations have been studied. A quite flexible alternative is the heterogenous Erdös-Rényi model  \citep{ouadah2020degree}.
\begin{definition}[Definition 2.]\textbf{[{Heterogenous Erdös-Rényi \textbf{(HER)} graph}]} \label{def:ER}
Let $G=(V,E)$ be a random graph on $n$ vertices with adjacency matrix $A$ and let $\bs{P}=(p_{ij})_{i,j=1,...,n}$ be the symmetric $(n\times n)$ matrix of connection probabilities. Then, we call $G$ a \textit{heterogenous Erdös-Rényi graph} $\mathcal{G}(n,\bs{P})$, if the edges are realizations of stochastically independent Bernoulli random variables. That is, for $1\leq i < j \leq n$, we have $A_{ij}\sim \mathnormal{\textup{Bin}}(1,p_{ij})$ with $A_{ji}:=A_{ij}$, and $A_{ii}:=0$ for all $i$. We denote the resulting HER model class by
\begin{align*}
\mathcal{G}_{HER}(n)=\{\mathcal{G}(n,\bs{P}), \bs{P}=(p_{ij}), p_{ii} = 0, p_{ij}=p_{ji}, p_{ij}\in[0,1]\ \forall\ i,j\}.
\end{align*}
\end{definition} 

In a nutshell, the HER-model expands the classical ER-model by offering the opportunity for individual link probabilities for each edge. This expansion is helpful for the modeling of more flexible scenarios, but also increases the complexity, e.g.~regarding parameter estimation, which might even become infeasible without further restrictions. Hence, we are interested in testing whether a parsimonious homogenous ER-model is already sufficient to model the underlying data or whether a HER-model achieves a significantly better fit. Consequently, having observed a simple graph $G$ of size $n$, we consider the testing problem
\begin{align}
\nullhyp~G\in\mathcal{G}_{ER}(n) \qquad \text{vs.} \qquad \alterhyp~G\in\mathcal{G}_{HER}(n) \backslash \mathcal{G}_{ER}(n). 
\label{testhyp}
\end{align}

Note that $\mathcal{G}_{ER}(n)\subset \mathcal{G}_{HER}(n)$ and, although $\mathcal{G}_{HER}(n)$ can be decomposed into disjoint sets, that is, in $\mathcal{G}_{ER}(n)$ and $\mathcal{G}_{HER}(n) \backslash \mathcal{G}_{ER}(n)$ as above, it will not be possible to consistently detect arbitrary alternatives, when testing $H_0$ against $H_1$. This is because a heterogenous ER model has to deviate \textit{sufficiently enough} from the homogenous model to be able to detect it, since we only rely on a single network observation in the goodness-of-fit context.

\subsection{Graph Functionals}\label{sec:GF}
In our work, we use the concept of graph functionals for the development of a general and rich class of goodness-of-fit tests. In this subsection, we gather classical existing theory for graph functionals and discuss extensions suitable for the derivation of general asymptotic theory for these goodness-of-fit tests.

\subsubsection{General Concept}\label{sec:GC}
Graph functionals cover a wide range of network statistics, which might be favorable for goodness of fit testing as they can capture various structural patterns of the rather complex nature of network data.
\begin{definition}[Definition 3.]\textbf{[Graph isomorphism and graph functional]} \label{def:GF}
	Let $G$ and $H$ be two simple graphs. An isomorphism of graph $G$ onto $H$ is a bijection $\varphi: V(G) \to V(H)$ such that any two vertices $v_1,\,v_2\in V(G)$ are adjacent in $G$ if and only if the vertices $\varphi(v_1),\,\varphi(v_2)\in V(H)$ are adjacent in $H$. If there exist an isomorphism between $G$ and $H$, we call $G$ and $H$ isomorphic. Furthermore, we denote by $\text{iso}_n(H)$ the set of isomorphisms on $n \geq n(H)$ vertices. A real-valued random variable $X_n=X_n(G)$ is called a \textit{graph functional} if it only depends on the isomorphism type of a graph $G$ of size $n$, i.e.~if $X_n(G)=X_n(H)$ holds for all $G$ and $H$ that are isomorphic.  
\end{definition} 
In other words, a graph functional is a function that does not hinge on the vertex labels. Although our proposed procedure in this paper is applicable to any kind of graph functional, we apply two special use cases throughout for illustration purposes.

\begin{example}[\textit{Example 1.}]\textbf{[Degree Variance]}
The \textit{degree variance} $V_n$ of a (simple) graph $G=(V,E)$ is defined as 
\begin{align*}
V_n := \frac{1}{n} \sum_{i=1}^n (D_i - \overline{D}_n)^2, 
\end{align*}
where $D_i$ is the degree of vertex $v_i$ and $\overline{D}_n=\frac{1}{n}\sum_{i=1}^n D_i$ denotes the average degree of the underlying network $G$. Note that the individual degrees $D_i$ do actually hinge on the vertex labels. However, it is easy to see that the sum of (squared) degrees as well as the average degree do not change by relabeling the graph. Hence, the degree variance $V_n$ is invariant under isomorphism change and, consequently, it is a graph functional. The degree variance is quite popular as it is an intuitive metric that can handily be computed. It is also used for goodness-of-fit testing in \cite{ouadah2020degree} and particularly serves as a heterogeneity index of a graph as pointed out by \cite{snijders1981degree}. In \cite{flossdorf2021change}, it is further investigated with the result that $V_n$ performs reasonable in capturing local structures of the graph (e.g.~centraility traits), but has weaknesses when it comes to capturing global characteristics like the overall amount of links.  
\end{example}

\begin{example}[\textit{Example 2}]\textbf{[(Centered) Subgraph counts]}\label{ex:Counts}
Let $G$ be a graph on $n$ vertices, and $H$ be another graph with $n(H)\leq n$. Then, the \textit{raw subgraph count} of $H$ in $G$ is defined by
	\begin{align}\label{Tn_def1}
	T_n(H) := \sum_{\widetilde{H} \in \text{iso}_n(H)} \prod_{e\in E(\widetilde{H})} \mathbbm{1}\{e\in E(G) \},
	\end{align}
The \textit{centered subgraph count} of $H$ in $G$ is defined by
	\begin{align}\label{Sn_def1}
	S_n(H) := \sum_{\widetilde{H} \in \text{iso}_n(H)} \prod_{e\in E(\widetilde{H})} \Big(\mathbbm{1}\{e\in E(G) \} - \mathbb{P}(e\in E(G))\Big). 
	\end{align}
	By construction, $S_n(H)$ is a random variable with expectation 0 for models with independently formed edges, as $\mathbb{P}(e\in E(G))=\mathbb{E}(\mathbbm{1}\{e\in E(G) \})$.

Note that in a relabeled graph, the subgraph of interest may be existent between different node combinations than in the observed graph. However, the sum of this type of subgraph stays the same over the whole network which is why those statistics also belong to the class of graph functionals. Subgraph counts are popular metrics for characterizing networks and are used for various inferential approaches \citep{maugis2020testing, gao2017testing, bhattacharyya2015subsampling}. However, these methods typically make use of subgraph counts having a specified simple shape and small order which makes them quite restrictive in their practical application. However, already for simple subgraph structures, the computational effort quickly becomes large in comparison e.g.~to the degree variance. Their usage as a graph functional, however, enables us to derive asymptotics in a unified way for flexible shapes and orders as we will show throughout the paper. 
\end{example}

\subsubsection{Unified Representation}
The key foundation of our unified procedure lies in a fundamental relationship between \textit{centered} subgraph counts and graph functionals: Any graph functional can be written as a linear combination of (possibly infinitely many) centered subgraph counts $S_n(H_i)$, $i=1,2,...$\,. This expansion enables the unified derivation of the asymptotic distributions of graph functionals $X_n$.
This method is referred to as the method of higher projections.
\begin{proposition}[Janson et al, 2011, Lemma 6.42] \label{prop:subgraph_decomposition}
A graph functional $X_n$ of a $\mathcal{G}(n,p)$ graph has a unique decomposition into variables $S_n(H),~H\in\mathcal{H}$, where $\mathcal{H}$ is a family of subgraphs without isolated vertices:
	\begin{align*}
	X_n = \sum_{H\in \mathcal{H}} a_n(H) S_n(H). 
	\end{align*}
	The $a_n(H)$'s are real-valued coefficients depending on $n,p,H$. The terms in the sum are orthogonal, hence
	\begin{align*}
	\Var(X_n) = \sum_{H\in\mathcal{H}} a_n^2(H) \Var(S_n(H)).
	\end{align*}
	The coefficients $a_n(H)$ are uniquely determined by 
	\begin{align*}
	a_n(H) = \frac{\E(X_n  S_n(H))}{\E(S_n^2(H))}.
	\end{align*}
\end{proposition}
For a given sequence of connection probabilities $p=p(n)\ntoinfty p_0\in[0,1]$, we call $X_n$ \textsc{dominated by a} family of connected graphs $\mathcal{H}_0$, i.e. $|\mathcal{H}_0|<\infty$, if
	\begin{align*}
	\frac{\Var(X_n)}{\sum_{H\in \mathcal{H}_0} a_n^2(H) \Var(S_n(H))} = \frac{\sum_{H\in\mathcal{H}} a_n^2(H) \Var(S_n(H))}{\sum_{H\in \mathcal{H}_0} a_n^2(H) \Var(S_n(H))} \xrightarrow[n\to\infty]{} 1. 
	\end{align*} 
	
Note that the expectation of $X_n$ is represented solely by the coefficient $a_n(\emptyset)$, i.e.~$E(X_n)=a_n(\emptyset)$. This is because the count of null graphs is deterministic and given by $S_n(\emptyset) = 1$ and all other terms $S_n(H)$, $H\not=\emptyset$, are centered by construction. Being dominated by a family of connected graphs $\mathcal{H}_0$ means that asymptotically all the variance of the graph functional $X_n$ is explained by information contained in a finite set of connected subgraphs. In particular, any disconnected subgraph in $\mathcal{H}$ will asymptotically not contribute to the variance as we will later see for the example of the degree variance below.

\subsubsection{Asymptotics}
This representation of graph functionals as a linear combination of centered subgraph counts enables the unified derivation of asymptotics for the whole class of graph functionals as we can use asymptotic results for $S_n(H)$ for this task. For deriving these asymptotics, it is convenient to slightly reformulate $S_n(H)$. To this purpose, let the $m=\binom{n}{2}$ possible edges of a graph $G$ be associated with independent random variables $Y_1,...,Y_m$ such that $Y_i$ has a Bernoulli distribution, i.e.
\begin{align*}
Y_i \sim \text{Bin}(1,p),\text{ where } p\in(0,1) \text{ for } i=1,...,m. 
\end{align*}
Thus, the edge set of a $\Gnp$ graph $G$ can be equivalently written as $E(G) \, = \, \{Y_1,...,Y_m\}$.
Alternatively to the representation in Example 2, subgraph counts can then be expressed in terms of the variables $Y_1,...,Y_m$. Replacing the indicators $\mathbbm{1}\{ e\in E(G) \}$ by the respective random variables $Y_i$ leads to the representations
\begin{align}\label{Tn_def2}
T_n(H) = \sum_{\tilde{H}\in \text{iso}_n(H)} \prod_{i:\,Y_i\in E(\widetilde{H})} Y_i. 
\end{align}
and
\begin{align}\label{Sn_def2}
S_n(H) = \sum_{\tilde{H}\in \text{iso}_n(H)} \prod_{i:\,Y_i\in E(\widetilde{H})} (Y_i - p) 
\end{align}
for the raw and centered subgraph counts, respectively. Asymptotic distributions of raw subgraph counts $T_n(H)$ in $\Gnp$ setups have been investigated in \cite{nowicki1988subgraph} and \cite{nowicki1989asymptotic}. Precisely, they showed that the subgraph counts of form $T_n(H)$ follow an asymptotic normal distribution if the underlying model is $\Gnp$. In our setup, however, we are interested in finding the asymptotic distributions for \textit{centered} subgraph counts $S_n(H)$ in order to get also asymptotic results for graph functionals via Proposition \ref{prop:subgraph_decomposition}. In this context, it is helpful that, for fixed $n$, we can calculate the expectation and variance of the statistics $S_n(H)$ as well as the covariance of two different counts $S_n(H)$ and $S_n(J)$. 
\begin{proposition}[Janson et al., 2011, Lemma 6.41]  \label{lemma:traits_of_sg_counts}
Let $H$ and $J$, $H,J\neq \emptyset$, be graphs without isolated vertices, let $|\text{aut}(H)|$ denote the cardinality of automorphisms of $H$, i.e. the number of isomorphisms from $H$ to itself, and $(x)_{y}=x(x-1)(x-2)\cdots(x-y+1)$ denotes the descending factorials.
Then, we have
	\begin{itemize}
		\item[(a)]  $\E(S_n(H)) =0$.
		\item[(b)]  $\Var(S_n(H)) = |\text{aut}(H)|(n)_{n(H)} (p(1-p))^{m(H)}$ 
		\item[(c)] If $H$ and $J$ are non-isomorphic, then $S_n(H)$ and $S_n(J)$ are orthogonal, that is, $\Cov(S_n(H), S_n(J)) = 0.$
	\end{itemize}
\end{proposition}
The closed and rather simple form of the variance of $S_n(H)$ stated in Proposition \ref{lemma:traits_of_sg_counts}(b) is a clear advantage opposed to raw subgraph count statistics $T_n(H)$. It results from the orthogonality of the different terms of the sum representing $S_n(H)$. Namely, we have
\begin{align*}
\Cov(Y_i-p, Y_j-p) = \E\Big((Y_i-p)\cdot(Y_j-p)\Big) = 0 \quad \forall \, i\neq j 
\end{align*}
due to the pairwise independence of the variables $Y_i$, $i=1,...,m$. The orthogonality of $S_n(H)$ and $S_n(J)$ stated in Proposition \ref{lemma:traits_of_sg_counts}(c) follows in a similar manner. This property is particularly helpful in deriving the joint distribution of \textit{multiple} subgraph counts. Under mild conditions, the statistics $S_n(H)$ with connected structures $H$ have an asymptotic normal distribution. 

\begin{proposition}[Janson et al., 2011, Theorem 6.43]\label{prop:normality_of_subgraph_counts}
Let $p=p(n)\ntoinfty p_0$ with $p_0\in[0,1]$. Then, for each unlabelled graph $H$ without isolated vertices with 
	\begin{align}\label{cond:Sparsity}
	np^{r(H)} \ntoinfty \infty, 
	\end{align}
	 with $r(H) = \max_{J \subseteq H} d(J)$, where $d(J) = m(J)/n(J)$, denotes the density of $J$, there exists a random variable $U(H)$ such that
		\begin{align}\label{Sn_CLT}
			n^{-n(H)/2}p^{-m(H)/2}S_{n}(H) \convD U(H).
		\end{align}
		 	 The convergence also holds jointly for any finite number of graphs $H$ satisfying the above condition $np^{r(H)} \ntoinfty \infty$. The limiting variables are determined by the following properties:
\begin{itemize}
	\item[(a)] If $H$ is connected and $m(H)> 0$, then
	\begin{align*}
		U(H) \sim \mathcal{N}(0, |\text{aut}(H)| (1 - p_0)^{m(H)})
	\end{align*}
\item [(b)] If $H_1,...,H_m$ are different (i.e. non-isomorphic) connected unlabelled graphs, then the random variables $U(H_1),...,U(H_m)$ are independent.
\end{itemize}
Furthermore, we have
\begin{align*}
	\E(U(H)^2) = |\text{aut}(H)| (1-p_0)^{m(H)} \quad \text{ for every }  H
\end{align*}
and for two different unlabelled graphs without isolated vertices $H_1$ and $H_2$, we have
\begin{align*}
	\E(U(H_1)U(H_2)) = 0.
\end{align*}
\end{proposition}
Note that Proposition \ref{prop:normality_of_subgraph_counts} generally allows for connection probabilities $p$ that remain fixed with increasing $n$, but also for sequences of connection probabilities $p=p(n)$ converging to zero at a certain rate. In this context, for a given subgraph $H$, the additional condition \eqref{cond:Sparsity} restricts the possible rates for $p(n)$. Precisely, this condition can be seen as a sparsity assumption and prevents $p(n)$ converging to 0 too fast.

From Proposition \ref{prop:subgraph_decomposition}, and the normality of the statistics $S_n(H)$ as stated in Proposition \ref{prop:normality_of_subgraph_counts}, we can obtain results on the asymptotic normality of graph functionals $X_n$. 
\begin{theorem}[Janson et al., 2011, Theorem 6.49] \label{thm:as_normality} 
	Let $X_n$ be a graph functional of $\mathcal{G}(n,p)$ with $p=p(n)\ntoinfty p_0 \in [0,1]$. Suppose $X_n$ is dominated by a family of connected graphs $\mathcal{H}_0$ such that, for all $H\in\mathcal{H}_0$ and $np^{r(H)} \ntoinfty \infty$ with $r(H) = \max_{J \subseteq H} d(J)$, where $d(J) = m(J)/n(J)$ denotes the density of $J$, the coefficients
	\begin{align*} 
	b(H) = \sup_n \frac{n^{n(H)/2} p^{m(H)/2} a_n(H) }{\sqrt{\Var(X_n)}}
	\end{align*}
	are finite and satisfy
	\begin{align*}
	\sum_{H\in \mathcal{H}_0} b(H)^2 |\text{aut}(H)| < \infty. 
	\end{align*}
	Then, as $n\rightarrow\infty$, it holds
	\begin{align*}
	\frac{X_n -\E(X_n)}{\sqrt{\Var(X_n)}} \xrightarrow{~d~} \mathcal{N}(0,1).
	\end{align*}
	In particular, there exists a sequence $(c_n)_{n\in\mathbb{N}}$ such that $c_n^2\Var(X_n)\rightarrow V^2>0$ leading to $c_n(X_n -\E(X_n)) \xrightarrow{~d~} \mathcal{N}(0,V^2)$.
\end{theorem}
The general representation of graph functionals as a linear combination of (possibly infinite) centered subgraph counts in Proposition \ref{prop:subgraph_decomposition} and the limiting distributions derived in Theorem \ref{thm:as_normality} form together a powerful tool that enables the flexible construction of a wide range of test statistics for goodness-of-fit testing for simple random graphs.

\section{Bootstrap theory for graph functionals}\label{sec:boots}
While knowledge of the presented limiting distributions for the class of graph functionals generally allows the construction and implementation of a powerful testing procedure, such asymptotic tests might have issues when it comes to small sample sizes, i.e. small network sizes $n$. Additionally, the derivation of the centered subgraph representation following Proposition \ref{prop:subgraph_decomposition} is not trivial and can be tedious. In order to provide a more flexible solution and to gain finite sample improvements, we propose to use bootstrapping to approximate the distribution of the test under the null. 

\subsection{Bootstrap Scheme}\label{sec:BootstrapScheme}
Having observed a simple random graph $G$, the bootstrap algorithm to estimate the distribution of a graph functional $X_n$ under the null of an ER-graph is defined as follows:
\begin{enumerate}
	\item[Step 1.] Estimate the connection probability $p$ by $\widehat{p} =\binom{n}{2}^{-1} \sum\limits_{1\leq i<j\leq n} A_{ij}$.
	\item[Step 2.] Conditional on $\bs{A} = (A_{ij})_{1\leq i,j\leq n}$, generate an ER graph $G^*$ by drawing a symmetric adjacency matrix $\bs{A}^*  = (A^*_{ij})_{1\leq i,j\leq n}$ from $\mathcal{G}(n,\hat{p})$. That is, conditional on $\bs{A} = (A_{ij})_{1\leq i,j\leq n}$, we $A_{ij}^*\sim \mathnormal{\textup{Bin}}(1,\widehat p)$ for $1\leq i < j \leq n$ with $A_{ji}^*:=A_{ij}^*$, and $A_{ii}^*:=0$ for all $i$.
	\item[Step 3.] Calculate the bootstrap graph functional $X_n^*=X_n(G^*)$.
	\item[Step 4.] Repeat Steps 2 and 3 $B$ times, where $B$ is large, to obtain $X_n^{*(b)}$, $b=1,...,B$. 
	\item[Step 5.] Approximate the distribution of graph functional $X_n$ by the empirical distribution of the bootstrap graph functionals $X_n^{*(1)},\ldots,X_n^{*(B)}$ (percentile bootstrap) or, alternatively, the distribution of the centered graph functional $X_n-\E(X_n)$ by the empirical distribution of the centered bootstrap graph functionals $X_n^{*(1)}-\E^*(X_n^*),\ldots,X_n^{*(B)}-\E^*(X_n^*)$ (Hall bootstrap), where $\E^*(\cdot)$ denotes the bootstrap expectation conditional on the original network.
\end{enumerate}

\subsection{Bootstrap Theory}\label{sec:BootstrapTheory}
In the following theorem, we provide asymptotic theory for the bootstrap procedure and prove its consistency in the framework of Theorem \ref{thm:as_normality}.  

\begin{theorem}\label{th:boot}
Suppose $G\in\mathcal{G}_{HER}$ with mean connectivity $p_{\text{mean}}=\binom{n}{2}^{-1} \sum_{1\leq i<j\leq n} p_{ij}$ such that $p_{\text{mean}}=p_{\text{mean}}(n)\ntoinfty p_0 \in [0,1]$. Further, let $X_n$ be a graph functional $X_n=X_n(G)$ computed from $G$. Suppose that, under the null hypothesis of a homogeneous ER graph, i.e.~when $G\in\mathcal{G}_{ER}\subset \mathcal{G}_{HER}$, we have the representation $X_n = \sum_{H\in \mathcal{H}} a_n(H) S_n(H)$ with $a_n(H)=a_n(H,p)$ that is dominated by a family of connected graphs $\mathcal{H}_0$. For all $H\in\mathcal{H}_0$, we suppose that $np_{\text{mean}}^{r(H)} \ntoinfty \infty$ with $r(H) = \max_{J \subseteq H} d(J)$, where $d(J) = m(J)/n(J)$ denotes the density of $J$, the coefficients
\begin{align}\label{boot_assumption1} 
b_{mean}(H) = \sup_n \frac{n^{n(H)/2} p_{\text{mean}}^{m(H)/2} a_n(H) }{\sqrt{\Var(X_n)}}
\end{align}
are finite and satisfy
\begin{align}\label{boot_assumption2}
\sum_{H\in \mathcal{H}_0} b_{mean}^2(H) |\text{aut}(H)| < \infty.
\end{align}
Further, let $X_n^*=X_n(G^*)$ denote the bootstrap version of the graph functional $X_n$ of $G^*$ generated according to the bootstrap scheme in Section \ref{sec:BootstrapScheme}. 
Then, we have $X_n^* = \sum_{H\in \mathcal{H}} a_n^*(H) S_n^*(H)$ with $a_n^*(H)=a_n(H,\widehat p)$ and, as $n\rightarrow\infty$, it holds
\begin{align*}
\frac{X_n^* -\E^*(X_n^*)}{\sqrt{\Var^*(X_n^*)}} \xrightarrow{~d~} \mathcal{N}(0,1)	\quad \text{in probability.}
\end{align*}
Additionally, with $(c_n)_{n\in\mathbb{N}}$ and $V^2$ as defined in Theorem \ref{thm:as_normality}, 
we have $c_n^2\Var^*(X_n^*)\rightarrow V^2>0$ in probability.
\end{theorem}

As a direct consequence of Theorems \ref{thm:as_normality} and \ref{th:boot}, under the stated assumptions, we get bootstrap consistency for the bootstrap procedure from Section \ref{sec:BootstrapScheme} in the sense that
\begin{align}\label{bootstrap_consistency_result}
\sup_{x\in\mathbb{R}}\left|P_{H_0}(c_n(X_n -\E(X_n))),P^*(c_n(X_n^* -\E^*(X_n^*)))\right|\rightarrow 0
\end{align}
and $c_n^2(\Var(X_n)-\Var^*(X_n^*))\rightarrow 0$ in probability, respectively, where $P_{H_0}$ denotes the probability under the null hypothesis of a homogeneous ER graph.
The bootstrap consistency result for the parametric bootstrap procedure in \eqref{bootstrap_consistency_result} enables the possibility to construct goodness-of-fit tests based on graph functionals not only in an asymptotic but also in a bootstrap manner. This has the great advantage that we do not rely anymore on finding the explicit centered subgraph representation of a graph functional as in Proposition \ref{prop:subgraph_decomposition} leading to a handily applicable class of goodness-of-fit procedures.

\section{Goodness-of-fit Testing for Erdös-Renyi Models}\label{sec:testing} 
In this section, we formulate our unified approach for goodness-of-fit testing for ER graphs. Precisely, we derive asymptotic level-$\alpha$ tests making use of the general asymptotic theory from Section \ref{sec:theory} and corresponding bootstrap tests based on the theory from Section \ref{sec:boots}. Additionally, we give further insights on the general use of centered and raw subgraph counts for goodness-of-fit testing. Furthermore, we investigate the sensitivity of several test statistics based on subgraph counts for certain SBM alternatives.

\subsection{A deep example: Degree variance goodness-of-fit testing}\label{sec:deg}
As our derived theory enables the construction of novel goodness-of-fit tests for a whole class of network metrics, let us first give a concrete example from the literature for our procedure. We consider the already mentioned degree variance statistic $V_n$ for which a goodness-of-fit test has been proposed by \cite{ouadah2020degree}. Eventually, we will see that - although using a different concept of proof - we come to the same result for our asymptotic version of the test which underlines the unified characteristic of our approach that contains this test as a special case.

\begin{figure}
	\centering
	\begin{framed}
		$H_1:$
		\begin{tikzpicture}
		\draw (0,0) -- (1,1);
		\fill[] (1,1) circle (1mm);
		\fill[] (0,0) circle (1mm);
		\end{tikzpicture} 
		\hspace{2cm} $H_2:$ \begin{tikzpicture}
		\draw (-1,1) -- (0,0) -- (1,1);
		\fill[] (-1,1) circle (1mm);
		\fill[] (0,0) circle (1mm);
		\fill[] (1,1) circle (1mm);		
		\end{tikzpicture}
		\hspace{2cm}  $H_3:$
		\begin{tikzpicture}
		\draw (-1,1) -- (0,0);
		\draw (1.5,1) -- (0.5,0);
		\fill[] (-1,1) circle (1mm);
		\fill[] (0,0) circle (1mm);
		\fill[] (1.5,1) circle (1mm);
		\fill[] (0.5,0) circle (1mm);		
		\end{tikzpicture}
	\end{framed}
	\caption{Subgraphs in the decomposition of the degree variance statistic.}
	\label{fig:subgraphs_in_decomposition}
\end{figure}
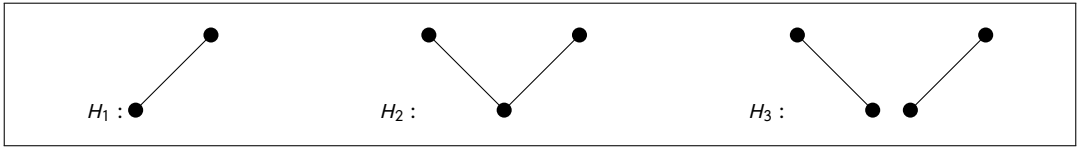

Firstly, we consider the asymptotic version of the test. In Section \ref{sec:GC}, we already argued that $V_n$ is a graph functional. In order to find its representation following Proposition \ref{prop:subgraph_decomposition}, we consider the centered version $V_n - \E(V_n)$, where $\E(V_n)=n^{-1}(n-1)(n-2)p(1-p)$, and use the Hoeffding decomposition \citep{hoeffding1992class, lee2019u} as in \cite{ouadah2020degree}. Then, under the null of a homogeneous ER graph, we get
\begin{align}
V_n  - \E(V_n) ~=~  &  \frac{2(n-2)}{n^2}(1-2p) \sum_{1\leq i < j\leq n}  \widetilde{A}_{ij}  +
\frac{2(n-4)}{n^2} \sum_{1\leq i < j < k\leq n} \left( \widetilde{A}_{ij}\widetilde{A}_{ik} +  \widetilde{A}_{ij}\widetilde{A}_{jk} +  \widetilde{A}_{ik}\widetilde{A}_{jk}\right) \nonumber    \\
& - \frac{8}{n^2} \sum_{1\leq i < j < k < l\leq n} \left( \widetilde{A}_{ij}\widetilde{A}_{kl} +
\widetilde{A}_{ik}\widetilde{A}_{jl} +
\widetilde{A}_{il}\widetilde{A}_{jk} \right)	\label{Vn1} \\
=:~ & V_n^{(A)} + V_n^{(B)} + V_n^{(C)}, \nonumber
\end{align}
where $\widetilde{A}_{ij} = A_{ij}-p$ are the centered entries of the adjacency matrix. As the true parameter $p$ is usually not known in practice, the consistent estimator $\widehat{p}$ can be used instead leading to the same asymptotics. This is guaranteed by Corollary \ref{cor:estimated_moments_Vn} below.
The components of the above decomposition are uncorrelated. This simplifies the derivation of the moments of $V_n$, especially of its variance leading to
\begin{align*}
\Var(V_n) = \Var\Big(V_n^{(A)}\Big) + \Var\Big(V_n^{(B)}\Big) +\Var\Big(V_n^{(C)}\Big). 
\end{align*}
Consequently, the variance $\Var(V_n)$ decomposes into three parts $\Var(V_n^{(A)})$, $\Var(V_n^{(B)})$ and $\Var(V_n^{(C)})$ that correspond to the  components that are depicted in Figure \ref{fig:subgraphs_in_decomposition} and can be calculated by 
\begin{align}\label{Vn2}
\Var(V_n) ~=~ &  \frac{2(n-1)(n-2)^2}{n^3}\,(1-2p)^2 \,p(1-p)  + \frac{2(n-1)(n-2)(n-4)^2}{n^3} \, \Big(p(1-p)\Big)^2 \\
& +  \frac{8(n-1)(n-2)(n-3)}{n^3} \, \Big(p(1-p)\Big)^2 .
\end{align}
Contrary to the presentations in \cite{ouadah2020degree}, we argue that it is not $V_n^{(A)}$, but the second part $V_n^{(B)}$, that makes the main contribution to the variance of $V_n$, if $np \ntoinfty \infty$. In particular, as $r(H)\geq 1$, the latter follows from
\begin{align}\label{Vn2a}
np^{r(H)} \ntoinfty \infty,
\end{align}
which is already assumed in Theorem \ref{thm:as_normality}. Intuitively, this allocation of the variances makes sense. The information contained in part $V_n^{(A)}$ is the centered number of edges in the observed network. This does not tell us much about the structure of the graph. However, part $V_n^{(B)}$ counts the number of edges that share a common node and are active -- it is a rescaled version of the centered count of two-stars (paths on three vertices with length two). We will also denote this subgraph as $P_3$ in the remainder of this paper. This part of the decomposition contains information on the local structure of the observed network. Part $V_n^{(C)}$ counts the number of pairs of \emph{disjoint} edges that are both active, again an information that does not tell us much about the global or local structure. Hence, it seems reasonable that $V_n^{(B)}$ contains most of the information of the network and for that explains most of the variance.
The decomposition into $V_n^{(A)}$, $V_n^{(B)}$ and $V_n^{(C)}$ is the linear combination of centered counts of the substructures $H_1$, $H_2$ and $H_3$, illustrated in Figure \ref{fig:subgraphs_in_decomposition}.
It can be embedded into Proposition \ref{prop:subgraph_decomposition} by setting
\begin{align}\label{Vn3}
&a_n(H_1) =  \frac{2(n-2)}{n^2}(1-2p),~  a_n(H_2) = \frac{2(n-4)}{n^2}, \text{ ~and~ }  a_n(H_3) = -\frac{8}{n^2}.\\
&\Var(S_n(H_1)) = p(1-p) \frac{n^2 - n}{2}, ~ \Var(S_n(H_2)) = (p(1-p))^2 \frac{(n-1)(n-2)n}{2}, \text{ ~and~ }\\ &\Var(S_n(H_3)) = (p(1-p))^2 \frac{(n-1)(n-2)(n-3)n}{8}.
\end{align}
Without loss of generality, we set $a_n(H_i)=0$ if $n<m(H_i)$ for any $i\in\{1,2,3\}$. Eventually, Theorem \ref{thm:as_normality} can be applied. For this, we need to check if its assumptions are fulfilled. Firstly, a dominating family of connected graphs is required. As $H_3$ is not connected and the variance contribution of $H_1$ is negligible, we set $\mathcal{H}_0 := \{H_2\}$. Then, we can verify that $\mathcal{H}_0$ is a dominating family of connected graphs for $V_n$ by checking the corresponding condition. Indeed, we have
	\begin{align}\label{Vn4}
	\frac{\Var(V_n)}{\sum\limits_{H\in \mathcal{H}} a_n^2(H)\, \Var(S_n(H))} = & \frac{ \frac{2(n-1)(n-2)^2}{n^3}p(1-p)(1+(n-6)p(1-p))}{\frac{4(n-4)^2}{n^4}\frac{n(n-1)(n-2)}{6} 3 (p(1-p))^2}
	= \frac{n-2 + (n^2-8n+12)}{(n^2-8n+16)} 
	\xrightarrow[n\to\infty]{} \,\, 1.
	\end{align}
	Consequently, the variance of $V_n$ is explained by the dominating family $\mathcal{H}_0$. Asymptotically it is solely driven by part $V_n^{(B)}$ of the Hoeffding decomposition. As a second step, we need to check whether the coefficient $b(H_2)$ is finite for any $n\geq 4$. This yields
	\begin{align}
	b(H_2) = & \sup_{n\geq 4} \frac{ n^{3/2}p\,\,\frac{2(n-4)}{n^2} }{ \sqrt{ \frac{2(n-1)(n-2)^2}{n^3} p(1-p)(1+(n-6)p(1-p)) } }	\label{Vn5}  \\
	\end{align}
	The finiteness of the supremum above is implied by
	\begin{align}
	& \lim_{n \to \infty}   \sqrt{ \frac{2(n^2-8n+16)n^2p}{(n-1)(n^2-4n+4)(1-p)(1+(n-6)p(1-p))} }	= \sqrt{\frac{2}{(1-p_0)^2}} <\infty  \quad \forall\, p_0\in[0,1),	\nonumber
	\end{align}
Furthermore, as the dominating family $\mathcal{H}_0$ is finite, this obviously leads to
	\begin{align}
	b^2(H_2) |\text{aut}(H_2)| = 6 \cdot \frac{2}{(1-p_0)^2} < \infty \quad \forall\, p_0\in[0,1).	\label{Vn6}
	\end{align}
Thus, from \eqref{Vn1} - \eqref{Vn6}, all conditions of Theorem \ref{thm:as_normality} are fulfilled yielding the asymptotic normality of $V_n - \E(V_n)$. That is, for $n\rightarrow\infty$, we have
\begin{align}
\frac{V_n -\E(V_n)}{\sqrt{\Var(V_n)}} \xrightarrow{~d~} \mathcal{N}(0,1).	\label{Vn_asymp}
\end{align}
In particular, for $c_n^2=(np^2(1-p)^2)^{-1}$, we have $c_n^2\Var(X_n)\rightarrow 2$ and $c_n(X_n -\E(X_n)) \xrightarrow{~d~} \mathcal{N}(0,2)$. As mentioned above, the true underlying parameter $p$ is usually not known in practice. In order to enable an implementable result for the shown asymptotics, we can make use of the following corollary.

\begin{corollary}\ \label{cor:estimated_moments_Vn} 
\begin{itemize}
	\item[(i)] Let $G\in\mathcal{G}_{ER}$ be a homogeneous ER graph $\mathcal{G}(n,p)$ with $p=p(n)\ntoinfty p_0 \in [0,1)$ such that $np^{r(H_2)} \ntoinfty \infty$. Then, the degree variance $V_n$ standardized by its estimated moments is asymptotically standard normal. That is, we have 
\begin{align*}
\frac{V_n -  \E_{\widehat{p}}(V_n)}{\sqrt{ \Var_{\widehat{p}}(V_n)}} \xrightarrow{~d~} \mathcal{N}(0,1), 	
\end{align*}
where $\E_{\widehat{p}}(V_n) = n^{-1}(n-1)(n-2)\widehat{p}(1-\widehat{p}) $ and $\Var_{\widehat{p}}(V_n) = n^{-3}2(n-1)(n-2)^2 \widehat{p}(1-\widehat{p})(1 + (n-6)\widehat{p}(1-\widehat{p}))$.
	\item[(ii)] Let $G\in\mathcal{G}_{HER}$ be a heterogeneous ER graph $\mathcal{G}(n,P)$ with mean connectivity $p_{\text{mean}}=\binom{n}{2}^{-1} \sum_{1\leq i<j\leq n} p_{ij}$ such that $p_{\text{mean}}=p_{\text{mean}}(n)\ntoinfty p_0 \in [0,1)$ with $np_{\text{mean}}^{r(H_2)} \ntoinfty \infty$. Then, the bootstrap degree variance $V_n^*$ standardized by its estimated bootstrap moments is asymptotically standard normal. That is, we have 
\begin{align*}
\frac{V_n^* - \E_{\widehat{p}^*}(V_n^*)}{\sqrt{\Var_{\widehat{p}^*}(V_n^*)}} \xrightarrow{~d~} \mathcal{N}(0,1), 	
\end{align*}
where $\E_{\widehat{p}^*}(V_n^*) = n^{-1}(n-1)(n-2)\widehat{p}^*(1-\widehat{p}^*) $ and $\Var_{\widehat{p}^*}(V_n^*) = n^{-3}2(n-1)(n-2)^2 \widehat{p}^*(1-\widehat{p}^*)(1 + (n-6)\widehat{p}^*(1-\widehat{p}^*))$.
\end{itemize}
\end{corollary}

The proof of Corollary \ref{cor:estimated_moments_Vn} can be found in Appendix A. An implementable asymptotic level$-\alpha$ test is thus given by the decision rule
\begin{align*}
\phi_{\alpha}(V_n) = \begin{cases}
1, & \text{ if } \left|\frac{V_n -  \E_{\hat{p}}(V_n)}{\sqrt{ \Var_{\hat{p}}(V_n)}} \right| > z_{1-\nicefrac{\alpha}{2}}, \\
0, & \text{ else,}
\end{cases}
\end{align*}
where $z_{1-\nicefrac{\alpha}{2}}$ is the $(1-\nicefrac{\alpha}{2})$ quantile of the standard normal distribution. Following the procedure explained in Section \ref{sec:boots}, a bootstrap version of the asymptotic level$-\alpha$ test is thus given by
\begin{align*}
\phi_{\alpha}^*(V_n) = \begin{cases}
1, & \text{ if } \left|\frac{V_n - \E_{\hat{p}}(V_n)}{\sqrt{\Var_{\hat{p}}(V_n)}} \right| > z^*_{1-\nicefrac{\alpha}{2}}, \\
0, & \text{ else,}
\end{cases}
\end{align*}
where $z^*_{1-\nicefrac{\alpha}{2}}$ is the $(1-\nicefrac{\alpha}{2})$ quantile of the empirical distribution of the resampled statistic $\frac{V_n^* -\E_{\widehat{p}^*}(V_n^*)}{\sqrt{\Var_{\widehat{p}^*}(V_n^*)}}$. An alternative bootstrap test of asymptotic level$-\alpha$ is given by 
\begin{align*}
\widetilde \phi_{\alpha}^*(V_n) = \begin{cases}
1, & \text{ if } \left|V_n - \E_{\hat{p}}(V_n) \right| > z^*_{1-\nicefrac{\alpha}{2}}, \\
0, & \text{ else,}
\end{cases}
\end{align*}
where $\widetilde z^*_{1-\nicefrac{\alpha}{2}}$ is the $(1-\nicefrac{\alpha}{2})$ quantile of the empirical distribution of the resampled statistic $V_n^* -\E_{\widehat{p}^*}(V_n^*)$.

\subsection{Goodness-of-fit testing based on subgraph counts}
The studied example underlines the applicability of our proposed class of tests both for the asymptotic as well as for the bootstrap version. Generally, this procedure can be performed for an arbitrary graph functional fulfilling the assumptions of Theorem \ref{thm:as_normality}. Thus, we now move on to investigate further traits of test statistics that are based on subgraph counts. In this context, we especially give a general result of Corollary \ref{cor:estimated_moments_Vn} for arbitrary shapes of $S_n(H)$. Additionally, we study to which extent our derived theory can also be applied to raw subgraph counts of shape $T_n(H)$.  
 
\subsubsection{Centered Subgraph Counts}\label{sec:Moments}
For general statistics $S_n(H)$ that are the foundation of our unified testing procedure, the construction of asymptotic tests is analogous to the example of the degree variance considered in detail in Section \ref{sec:deg} as the theory gathered in Section \ref{sec:theory} can be applied. For the final formulation of an implementable test, we can extend Corollary \ref{cor:estimated_moments_Vn} to a more generalized version.

\begin{theorem} \label{rem:estmoments}
Let $G\in\mathcal{G}_{ER}$ be a homogeneous ER graph $\mathcal{G}(n,p)$ with $p=p(n)\ntoinfty p_0 \in [0,1)$. Further, let  $H$ be a connected subgraph and assume $np^{r(H)} \ntoinfty \infty$. We define
	\begin{align*}
	\widehat{S}_n(H) := \sum_{\widetilde{H}\in \text{iso}_n(H)} \prod_{e\in E(\widetilde{H})} (\mathbbm{1}\{e\in  E(G)\}-\widehat{p}),
	\end{align*}
	where $\hat{p} =\binom{n}{2}^{-1} \sum\limits_{1\leq i<j\leq n} A_{ij}$ such that $\widehat p-p=O_P(\frac{p}{n})$. Then, we have
	\begin{align*}
	\frac{ \widehat{S}_n(H)}{ \sqrt{|\text{aut}(H)|(n)_{n(H)} (\widehat p(1-\widehat p))^{m(H)} } } \convD \mathcal{N}(0,1). 
	\end{align*}
That is, standardizing statistics $S_n(H)$ with their estimated moments does not change the asymptotic distribution.
\end{theorem}

Thus, an asymptotic level-$\alpha$ tests for hypotheses \eqref{testhyp} based on the centered subgraph counts $S_n(H)$ 
are given by decision rules
\begin{align*}
\phi_{\alpha}(S_n(H)) := \begin{cases}
1, & \text{if~ } \left|\frac{\widehat{S}_n(H)}{\sqrt{|\text{aut}(H)|(n)_{n(H)} (\widehat p(1-\widehat p))^{m(H)} }}\right| > z_{1-\nicefrac{\alpha}{2}}, \\
0, & \text{else},
\end{cases}
\end{align*}
where $z_{1-\nicefrac{\alpha}{2}}$ is the $(1-\nicefrac{\alpha}{2})$ quantile of the standard normal distribution. The proof of Theorem \ref{rem:estmoments} can be found in Appendix A. A bootstrap version can be easily implemened with the usage of Theorem \ref{th:boot} analogously to the ones at the end of Section \ref{sec:deg} above.

\subsubsection{Raw Subgraph Counts} \label{sec:raw_subgraph_counts}
Although we mostly focus on the behaviour of centered subgraph counts $S_n(H)$, it is a logical question whether the derived theory can be extended to the more intuitive concept of raw subgraph counts $T_n(H)$. For illustration, we will focus on $T_n(C_3)$, i.e.~the count of triangles.
	
\cite{nowicki1988subgraph} and \cite{nowicki1989asymptotic} prove that subgraph counts of shape $T_n(H)$ follow an asymptotic normal distribution if the underlying model is $\Gnp$. The crux in their derivation is an approximation based on conditional expectations to avoid the heavy calculations for the variance which result from the correlated terms in the raw subgraph counts.
The proof relies on advanced techniques for U-statistics.

For $T_n(C_3)$, Theorem 4 from \cite{nowicki1988subgraph} gives 
\begin{align} \label{eq:TnC3_approx}
\frac{T_n(C_3) - \binom{n}{3}p^3}{(n-2)\sqrt{\binom{n}{2}p^5(1-p)}} \convD \mathcal{N}(0,1).
\end{align} 
Since $T_n(C_3)$ is a graph functional, we can also apply the method of higher projections to obtain this result. It can easily be shown that 
\begin{align} \label{eq:TnC3_decomposition}
	T_n(C_3) = S_n(C_3) +  p S_n(P_3) + p^2  (n - 2)  S_n(P_2) + p^3  |aut(C_3)|.
\end{align} 
The closed form variance of $T_n(C_3)$ can thus handily be derived as 
\begin{align} \label{eq:TnC3_variance}
	\Var\left(T_n(C_3)\right) = \binom{n}{3}\left(p(1-p)\right)^3 + 3\binom{n}{3}p^2\left(p(1-p)\right)^2 + \binom{n}{2}(n-2)^2p^5(1-p).
\end{align}
It can be shown that the variance is dominated by the contribution of $S_n(P_2)$, which is exactly the term used for standardization in \eqref{eq:TnC3_approx}. This directly proves \eqref{eq:TnC3_approx} by applying Proposition \ref{prop:subgraph_decomposition} and Theorem \ref{thm:as_normality} with dominating family $\mathcal{H}_0=\{P_2\}$.

In the manner of the previous subsection, one could try to construct a test for hypotheses \eqref{testhyp} based on the count of triangles $T_n(C_3)$:
In order to be able to use this representation for a test of hypotheses \eqref{testhyp}, we need to plug in estimates of the expectation and variance based on $\hat{p}$. However, in this case the standardized test statistic, given as
\begin{align} \label{eq:tn_est}
	\widetilde{T} = \frac{T_n(C_3) - \binom{n}{3}\widehat{p}^3}{(n-2)\sqrt{\binom{n}{2}\widehat{p}^5(1-\widehat{p})}}
\end{align}
no longer follows a standard normal distribution. This is because we obtain
\begin{align}
	\frac{T_n(C_3) - \E_{\widehat{p}}(T_n(C_3))}{\sqrt{\Var_{\widehat{p}}(T_n(C_3))}} & = \frac{T_n(C_3) - \E_{\widehat{p}}(T_n(C_3)) - \E_p(T_n(C_3)) + \E_p(T_n(C_3))}{\sqrt{\Var_{\widehat{p}}(T_n(C_3))}} \frac{\sqrt{\Var_{p}(T_n(C_3))}}{\sqrt{\Var_{p}(T_n(C_3))}} \\
	& = \underbrace{\frac{\sqrt{\Var_{p}(T_n(C_3))}}{\sqrt{\Var_{\widehat{p}}(T_n(C_3))}}}_{\xrightarrow{n\to\infty} 1} \left( \underbrace{\frac{T_n(C_3) - \E_p(T_n(C_3))}{\sqrt{\Var_{p}(T_n(C_3))}}}_{\convD \mathcal{N}(0,1)} + \underbrace{\frac{\E_p(T_n(C_3)) - \E_{\widehat{p}}(T_n(C_3))}{\sqrt{\Var_{p}(T_n(C_3)n)}}}_{=: T} \right)
\end{align}
For the second term in brackets, one can also show with an application of the delta method \citep{van2000asymptotic} that
\begin{align}
T = \frac{\E_p(T_n(C_3)) - \E_{\widehat{p}}(T_n(C_3))}{\sqrt{\Var_{p}(T_n(C_3))}} \convD \mathcal{N}(0, 1).
\end{align}
Thus, the limiting distribution of the potential test statistic changes if we plug in estimated moments.

Since $\widehat{p}^3$ is a graph functional itself and, consequently, also $T_n(C_3) - \binom{n}{3}\widehat{p}^3$, we again need to determine the dominating family, but now for the centered raw subgraph count $T_n(C_3) - \binom{n}{3}\widehat{p}^3$. Interestingly, the main variance contribution no longer comes from $S_n(P_2)$, but instead is explained by the other two terms in the decomposition \eqref{eq:TnC3_decomposition}, i.e.~$C_3$ and $P_3$. 

The following lemma derives the dominating family for $\widehat{p}^3$, the term used for centering in \eqref{eq:tn_est}.

\begin{lemma}[Dominating family for $\widehat{p}^3$]\label{lemma:phat}
	Let $\widehat{p} = \binom{n}{2}^{-1} \sum_{1\leq i < j \leq n} A_{ij}$ be an estimate for the parameter $p$ of an Erdös-Renyi graph. As $\hat{p}$ is a graph functional, it we can be expressed in terms of centered subgraph counts. That is, we have
	\begin{align} \label{eq:phat_subgraph}
		\widehat{p} = \binom{n}{2}^{-1} S_n(P_2) + p. 
	\end{align}
Obviously, the statistic $\widehat{p}^3$ has dominating family $\mathcal{H}_0=\{P_2\}$ and, by using the method of higher projections, we can write
	\begin{align}\label{eq:dom_fam}
		\widehat{p}^3 = 3\binom{n}{2}^{-1}p^2 S_n(P_2) + p^3 + \mathcal{O}_{\prob}(n^{-2}).
	\end{align}
\end{lemma}
From Lemma \ref{lemma:phat}, we can conclude that
\begin{align} \label{eq:phat3_approx}
	\binom{n}{3}\hat{p}^3 = (n-2)p^2S_n(P_2) + \binom{n}{3}p^3 + \mathcal{O}_{\prob}(n).
\end{align}
Combining \eqref{eq:phat3_approx} and \eqref{eq:TnC3_decomposition}, we get
\begin{align} \label{eq:Tn_centered}
T_n(C_3) - \binom{n}{3}\widehat{p}^3 = S_n(C_3) + p\cdot S_n(P_3) + \mathcal{O}_{\prob}(n)
\end{align}
and the leading term of the variance of $T_n(C_3) - \binom{n}{3}\widehat{p}^3$ becomes
\begin{align}\label{eq:var_stand}
\binom{n}{3} p^3(1-p)^3 + 3\binom{n}{3}p^3(1-p)^2,
\end{align}
which is the contribution to the variance are made by $C_3$ and $P_3$, while the effect of $P_2$ is canceled out through subtraction of $\widehat{p}^3$ from the raw subgraph count $T_n(C_3)$. The findings are summarized in the following Corollary \ref{cor:TnC3}. The obtained variance is exactly the one reported in Theorem 2.2 in \cite{gao2017testing} for the same test statistic. This shows how the unifying theory for the graph functionals pays off: It allows to directly derive limiting distributions without further knowledge about the statistic.


\begin{corollary}[Dominating family and asymptotic distribution of $\widetilde{T}$ in \eqref{eq:tn_est}] \label{cor:TnC3}
	Given the model $\Gnp$ with $p\in(0,1)$. The raw count of triangles centered with its estimated mean, that is,
	\begin{align}
		T_n(C_3) - \binom{n}{3} \widehat{p}^3,
	\end{align}
is a graph functional. It is dominated by $\mathcal{H}_0 = \{C_3, P_3\}$ where $a_n(C_3) = 1$ and $a_n(P_3) = p$. Theorem \ref{thm:as_normality} is applicable and yields
\begin{align}\label{eq:triangle_test_statistic}
\frac{T_n(C_3) - \binom{n}{3}\widehat{p}^3}{\sqrt{\binom{n}{3}\widehat{p}^3 (1-\widehat{p})^3 + 3\binom{n}{3}\widehat{p}^4(1-\widehat{p})^2}} \xrightarrow{d} \mathcal{N}(0, 1).
\end{align}
\end{corollary}

Thus, an asymptotic level-$\alpha$ tests for hypotheses \eqref{testhyp} based on the test statistic \eqref{eq:triangle_test_statistic} 
are given by decision rules 
\begin{align*}
	\phi_{\alpha}(\widehat{T}_n(C_3) ) := \begin{cases}
		1, & \text{if~ }\left\lvert\frac{T_n(C_3) - \binom{n}{3}\widehat{p}^3}{\sqrt{\binom{n}{3}\widehat{p}^3 (1-\widehat{p})^3 + 3\binom{n}{3}\widehat{p}^4(1-\widehat{p})^2}}\right\rvert > z_{1-\nicefrac{\alpha}{2}}, \\
		0, & \text{else},
	\end{cases}
\end{align*}
where $z_{1-\nicefrac{\alpha}{2}}$ is the $(1-\nicefrac{\alpha}{2})$ quantile of the standard normal distribution. 

Another option for a test of hypotheses \eqref{testhyp} is offered by the bootstrap methodology presented in Section \ref{sec:BootstrapTheory}. We can avoid having to standardize by applying the bootstrap scheme presented in Section \ref{sec:BootstrapScheme}. 
We can directly use $T_n(C_3)$ as the tests statistic and approximate its distribution by the resampling distribution of the statistics $T_n^*(C_3)$, $b=1,...,B$. A test for the null hypotheses $H_0$ in \eqref{testhyp} can be performed using the following decision rule 
\begin{align*}
	\phi_\alpha(T_n(C_3))^* := \begin{cases}
		1, & \text{ if } {T}_n(C_3) < q^*_{\nicefrac{\alpha}{2}} \text{ or } {T}_n(C_3) > q^*_{1-\nicefrac{\alpha}{2}}, \\
		0, & \text{ else,}
	\end{cases}
\end{align*}
\enlargethispage{2\baselineskip}
where $q^*_{\nicefrac{\alpha}{2}}$ and $q^*_{1-\nicefrac{\alpha}{2}}$ are the $\nicefrac{\alpha}{2}$ and the $(1-\nicefrac{\alpha}{2})$ quantile of the bootstrap distribution of the $T_n^{*}(C_3)$.

\subsection{Power Analysis for SBM alternatives}\label{sec:power}
The main advantage of our unified goodness-of-fit testing procedure is that it enables the user to use an arbitrary graph functional that fulfills the assumptions of Theorem \ref{thm:as_normality} for the construction of the test statistic without tediously deriving the necessary asymptotics for each single case. A further valuable aspect is that the class of graph functionals is quite broad, i.e.~there are a lot of different test statistics to choose from resulting in a large flexibility for various application fields. However, as this class is quite broad, it is crucial to choose an appropriate graph functional for a succesful application and implementation in the underlying situation. Obviously, each possible graph functional might be sensitive to different network information and thus might be more or less suitable for a particular data setup. This raises the question of which graph functional to use in which situation. Because general statements are obviously very difficult to derive, in the following, 
we give a concrete theory-driven analysis for a specific example based on triangles, denoted as $C_3$, and two-stars, denoted as $P_3$. Further examples are then investigated in an extensive simulation study in Section \ref{sec:Sim}. 

For now, we consider the popular case of Stochastic Block Models (SBM) that particularly plays an important role in modeling social network behaviours and patterns. It assumes that the set of $n$ nodes can be divided into $K$ blocks. Typically, nodes of the same block then have a higher probability to share an edge than nodes of different blocks. SBMs of size $n$ are thus low-parametrized special cases of $\mathcal{G}_{HER}(n)$ models. We restrict our following power analysis to SBMs with $K=2$ blocks, equal block sizes (i.e.~$\nicefrac{n}{2}$) and equal intra-group probabilities $p_{\text{intra}}$ for both blocks. The edge probability between the groups are denoted by $p_{\text{inter}}$ such that $p_{\text{intra}} \geq p_{\text{inter}}$. For $p_{\text{intra}} = p_{\text{inter}}$, the SBM becomes a $\mathcal{G}_{ER}(n)$ model. To rule out detection just based on the total number of edges, we generate models from both model classes such that their mean connectivity parameters $p_{\text{mean}}$ (which is allowed to depend on $n$) coincide. Note that, obviously, $p_{\text{mean}} = p$ for the $\mathcal{G}_{ER}(n)$ model class.


\subsubsection{Triangles $C_3$}
Let us first focus on the performance of $C_3$. Therefore, we need to have a closer look into the expectation of centered triangle counts for ER and SBM models. In general, that is for both model classes, $\E(S_n(C_3))$ can be decomposed as follows
\begin{align*}
\E(S_n(C_3))  = (1 - p)^3\E(I_n(C_3)) + (1 - p)^2(-p)\E(I_n(P_3)) + (1 - p)p^2\E(I_n(D_3)) + (-p)^3\E(I_n(E_3)),
\end{align*}
where the terms $I_n(H)$ denote counts of induced subgraphs. Furthermore, $D_3$ is a subgraph consisting of three nodes and one edge and $E_3$ a subgraph consisting of three nodes and no edges. Thus, $\E(S_n(C_3))$ can be divided into a linear combination of expected values of all possible subgraph shapes consisting of three nodes. The coefficients represent the probabilities for the occurence of the corresponding subgraph. In case of a homogenous ER-model, by construction, these terms can be handily determined to get
\begin{align*}
\E_\text{ER}(S_n(C_3)) & = (1 - p)^3\bin p^3 + (1 - p)^2(-p)3\bin (1 - p)p^2 + (1 - p)p^2 3\bin (1 - p)^2 p + \bin (1- p)^3(-p)^3  \\
&= (1 - p)^3\bin p^3 - p^3\bin (1 - p)^3 + (1 - p)^3p^3 3\bin - (1 - p)^3 p^3 3 \bin	\\
&= 0.
\end{align*}
In case of an SBM, the situation is more complicated. While the corresponding formulas for $\E_\text{SBM}(I_n(C_3))$ and $\E_\text{SBM}(I_n(E_3))$ are rather straightforward to derive, the derivation of the formulas for $\E_\text{SBM}(I_n(P_3))$ and $\E_\text{SBM}(I_n(D_3))$ are more complicated and can be found in the Appendix. Altogether, we obtain
\begin{align}\label{eq:SBM_C3}
\begin{split}
\E_\text{SBM}(S_n(C_3)) &= (1 - p_{\text{mean}})^3 \left[2 \binHalf \pra^3 + \left[\left(\bin - 2\binHalf \right) \pra \per^2\right] \right] \\
& + (1 - p_{\text{mean}})^2 (-p_{\text{mean}}) \left[6 \binHalf \pra^2 (1 - \pra) + 2n \binom{\nicefrac{n}{2}}{2} \pra \per (1 - \per) + n \binom{\nicefrac{n}{2}}{2} \per^2 (1 - \pra) \right] \\
& + (1 - p_{\text{mean}})p_{\text{mean}}^2 \left[6 \binHalf \pra (1 - \pra)^2 + 2n \binom{\nicefrac{n}{2}}{2} (1 - \pra) \per (1 - \per) + n \binom{\nicefrac{n}{2}}{2} \pra (1 - \per)^2 \right] \\
& + (-p_{\text{mean}})^3 \left[2 \binHalf (1 - \pra^3) + \left[\left(\bin - 2\binHalf \right) (1 -\pra) (1 - \per)^2\right]\right].
\end{split}
\end{align}
These results lead to the following useful Theorem.
\begin{theorem}\label{th:C3}
Consider a homogenous ER-model and a Stochastic Block model (SBM) with two blocks, equal block sizes and equal intra-group probability. Suppose that the mean connectivity of both networks is the same, that is $p$ for the $\mathcal{G}_{ER}$ model and $p_{\text{mean}}$ for the SBM model coincide. Then, for the centered subgraph count $S_n(C_3)$ and the raw subgraph count $T_n(C_3)$, respectively, we get
\begin{itemize}
\item[(a)] $\lim\limits_{n\rightarrow \infty} \E_{\text{ER}} (T_n(C_3)) - \E_{\text{SBM}}(T_n(C_3)) < 0, \text{ if } p_{\text{intra}} > p_{\text{inter}}$,
\item[(b)] $\lim\limits_{n\rightarrow \infty} \E_{\text{ER}} (S_n(C_3)) - \E_{\text{SBM}}(S_n(C_3)) < 0, \text{ if } p_{\text{intra}} > p_{\text{inter}}$.
\end{itemize} 
\end{theorem}   
The proof of both statements can be found in the Appendix. 
In the context of our work, we are more interested in statement (b) about the behaviour of the centered subgraph count $S_n(C_3)$. However, in both cases, the Theorem guarantees that the expectation for an ER-model differs (asymptotically) from the expectation computed from an SBM model. Thus, $C_3$ is sensitive to distinguishing between both models. However, to give a reliable answer for the suitability of $S_n(C_3)$ for goodness-of-fit testing, we need to evaluate to which extent this sensitivity exists. Before we analyze this in more details, let us first concentrate on similar derivations for subgraph counts based on $P_3$. 

\subsubsection{Two-stars $P_3$}
Similar to the derivation above for triangles $C_3$, we can derive a corresponding result also for two-stars. In the end, it yields $\E_\text{ER}(S_n(C_3)) = 0$ and
\begin{align}\label{eq:SBM_P3}
\begin{split}
&\E_\text{SBM}(S_n(P_3)) = 3(1 - p_{\text{mean}})^2 \left[2 \binHalf \pra^3 + \left[\left(\bin - 2\binHalf \right) \pra \per^2\right] \right] \\
& + \left[(1 - p_{\text{mean}})^2 + 2(-p_{\text{mean}})(1-p_{\text{mean}})\right] \left[6 \binHalf \pra^2 (1 - \pra) + 2n \binom{\nicefrac{n}{2}}{2} \pra \per (1 - \per) + n \binom{\nicefrac{n}{2}}{2} \per^2 (1 - \pra) \right] \\
& + \left[(-p_{\text{mean}})^2 + 2(-p_{\text{mean}})(1-p_{\text{mean}})\right] \left[6 \binHalf \pra (1 - \pra)^2 + 2n \binom{\nicefrac{n}{2}}{2} (1 - \pra) \per (1 - \per) + n \binom{\nicefrac{n}{2}}{2} \pra (1 - \per)^2 \right] \\
& + 3(-p_{\text{mean}})^2 \left[2 \binHalf (1 - \pra^3) + \left[\left(\bin - 2\binHalf \right) (1 -\pra) (1 - \per)^2\right]\right]
\end{split}
\end{align}
leading to the following result. 

\begin{theorem}\label{th:P3}
Consider a homogenous ER-model and a Stochastic Block model (SBM) with two blocks, equal block sizes and equal intra-group probability. Suppose that the mean connectivity of both networks is the same, that is $p$ for the $\mathcal{G}_{ER}$ model and $p_{\text{mean}}$ for the SBM model coincide. Then, for the centered subgraph count $S_n(C_3)$ and the raw subgraph count $T_n(C_3)$, respectively, we get
\begin{itemize}
\item[(a)] $\lim\limits_{n\rightarrow \infty} \E_{\text{ER}} (T_n(P_3)) - \E_{\text{SBM}}(T_n(P_3)) > 0, \text{ if } p_{\text{intra}} > p_{\text{inter}}$,
\item[(b)] $\lim\limits_{n\rightarrow \infty} \E_{\text{ER}} (S_n(P_3)) - \E_{\text{SBM}}(S_n(P_3)) > 0, \text{ if } p_{\text{intra}} > p_{\text{inter}}$.
\end{itemize} 
\end{theorem}   
The derivations and proofs are analogous to before and can be found in the Appendix. 

\subsubsection{Discussion}
\begin{figure}
\centering
\includegraphics[width=3.1in, keepaspectratio = TRUE]{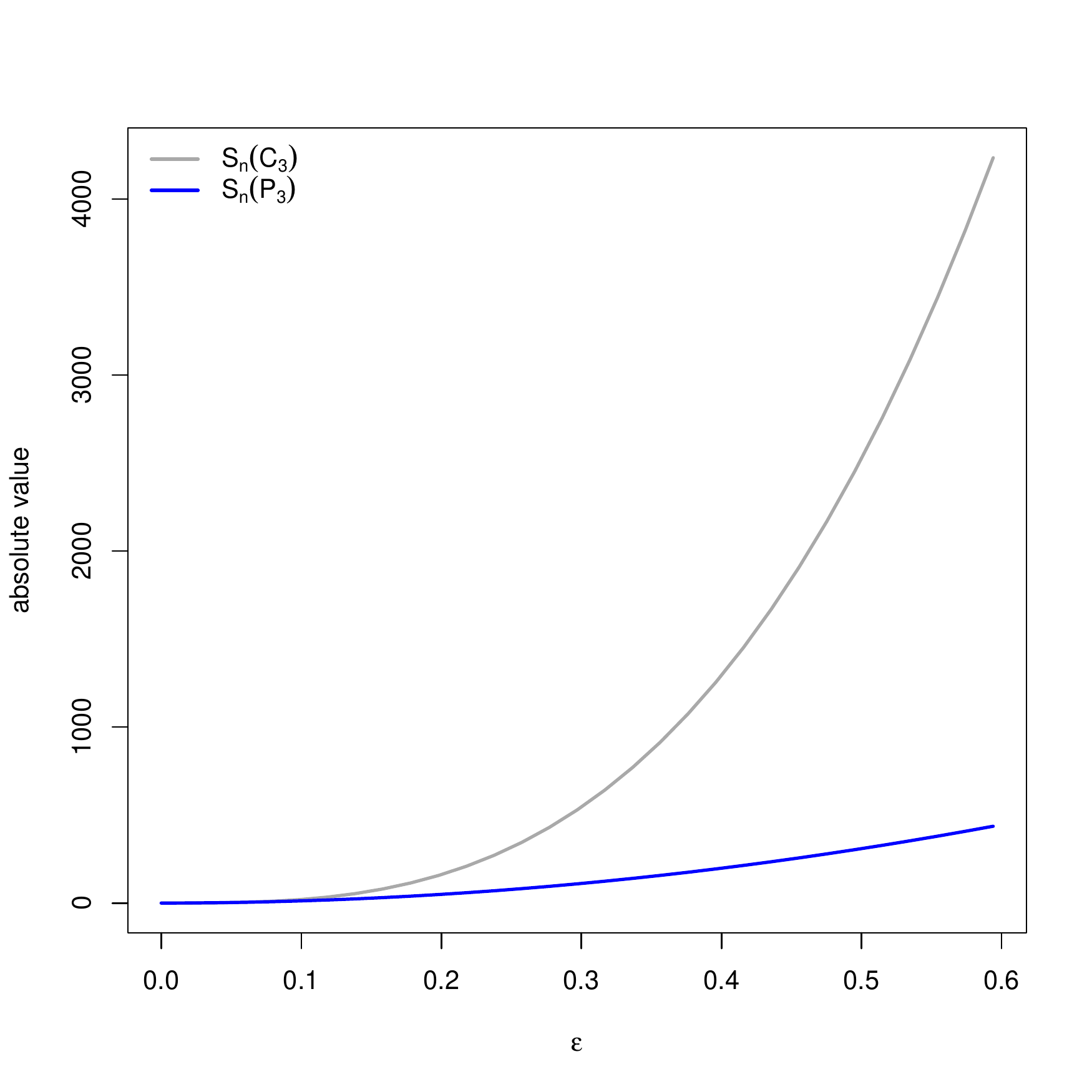}
\caption{Expected absolute values of centered subgraph counts in the SBM case with $p_{\text{mean}} = 0.3$ and $n = 100$.}\label{fig:power}
\end{figure}

An investigation of the presented results for $\E_\text{SBM}(S_n(P_3))$ and $\E_\text{SBM}(S_n(C_3))$ enables us 
to analyze to which extent the sensitivity for distinguishing SBMs from ER-models is pronounced for both subgraph counts. Note that the analysis of the expected value in the SBM case is sufficient in this context, since the expected value for ER-models is 0 as shown. In this context, Figure \ref{fig:power} illustrates the behaviour of both derived functions (absolute values) for different parameters $\epsilon = \pra - \per$. 
Since we are in the situation of centered subgraph counts, the expected value only depends on $\epsilon$. Particularly, the mean connectivity $p_{\text{mean}}$ does not have direct influence. It has only indirect influence on how large $\epsilon$ can be chosen, e.g. for $p_{\text{mean}} = 0.5$ there are more values of $\epsilon$ possible than for $p_{\text{mean}} = 0.1$ or $p_{\text{mean}} =0.9$ as $\pra, \per \in [0,1]$ is required. Hence, Figure \ref{fig:power} that is constructed for an $p_{\text{mean}} = 0.3$ and $n = 100$ stands representative for arbitrary mean connectivity parameter $p_{\text{mean}} \in [0,1]$. It is clearly visible that the absolute value of $\E(S_n(C_3))$ for SBMs is larger than $\E(S_n(P_3))$. This behaviour is even more pronounced for larger values of $\epsilon$. Consequently, the triangle structure is way more sensitive to detecting these types of SBMs. Intuitively, this makes sense as the condition for triangles are more restrictive than for two-stars, since all links between the three involved nodes have to exist. Obviously, this is easier to achieve in the dense blocks of a SBM than in an ER-graph due to $p_\text{intra} > p_{\text{mean}}$. These results are confirmed by the simulation study in Section \ref{sec:Sim}, where we also evaluate the behaviour for more flexible SBM setups with different parameters.

\section{Simulation Study}\label{sec:Sim}
To underline our findings and to gain further evidence for the proposed class of tests, we execute a simulation study for different parameter setups and various alternatives. We analyze the power of the tests and illustrate performance differences depending on the chosen graph functional.

\subsection{General Setting}
We investigate three different graph functionals for the construction of the test statistics. Namely, these are the degree variance statistic $V_n$, centered triangle counts $S_n(C_3)$ and centered two-star counts $S_n(P_3)$. For all three, we analyze their performances for the asymptotic version and for the boostrap version of the testing procedure. To investigate the power of these test statistics, we have to generate networks from the alternative, i.e from the $\GnP$ model. In order to be able to detect the alternative, the generated networks have to deviate from the homogeneous null model by an increasing amount with increasing heterogenity. To generate random graphs under the alternative, we set a mean connectivity $p_{\text{mean}}$ such that the null model is a $\mathcal{G}(n,p_{\text{mean}})$ model. This is to ensure that the power of the tests is not due to differences in the mean connectivity (thus differing degrees and mean degrees), but actually due to the rising heterogeneity of the connection probabilities. Hence, we make the requirement:
\begin{align*}
\binom{n}{2}^{-1} \sum_{1\leq i <j \leq n} p_{ij} = p_{\text{mean}}.
\end{align*} 
This preserves the mean connectivity over all scenarios and settings. Although this limits the possible alternatives, it is not a big constraint. Having an observed network $G$, we can directly calculate its mean connectivity $p_{\text{mean}}$. In order to study the performances in various network size and density conditions, we use different values of $n = \{16, 32, 64, 128\}$ and $p_{\text{mean}} = p_{\text{mean}}(n) = \left\{\frac{\log{n}}{n}, \frac{1}{\sqrt{n}}, \frac{\log{n}}{\sqrt{n}} \right\}$. Note that for these combinations of $n$ and $p_{\text{mean}}$ we have $\frac{\log{n}}{n} < \frac{1}{\sqrt{n}} < \frac{\log{n}}{\sqrt{n}}$.

As the $\GnP$ model class is very broad, we limit the simulation study to a few relevant scenarios for the alternative that serve as representatives for different possible ways the matrix $\Pmat$ can be set up. These include the popular use case of SBMs and covariate models. For the SBMs, we use two different setup versions including the usage of two blocks with equal block sizes as in our analysis of Section \ref{sec:power} and the extension to three blocks with random block sizes and varying intra-group probabilities. The exact setup of each alternative will be explained in the corresponding subsections below.

We calculate the power of the tests by the application of 1000 replications for each setting. For the bootstrap version, we use $B = 500$ bootstrap replications. We set the test level as $\alpha = 0.05$ for all tests.
  
\subsection{Performances}
With the consideration of all described parameter variations, we investigate the power of 72 different scenarios. The main results and important insights are presented in the following.

\subsubsection{Stochastic block models}
As mentioned above, we construct two different versions of SBMs. The first one is characterized by two blocks of equal block sizes and is constructed as follows. We assign weights to the nodes with
\begin{align*} 
\bs{w} = (w_1,...,w_n) = \bigg( \underbrace{-\frac{1}{2},...,-\frac{1}{2}}_{\nicefrac{n}{2} \text{ times}}, \underbrace{\frac{1}{2},...,\frac{1}{2}}_{\nicefrac{n}{2} \text{ times}}\bigg). 
\end{align*}
By premultiplying the weights with a factor $\lambda>0$, we can achieve different degrees of heterogeneity. 
Concretely, the weights are transformed into connection probabilities $p_{ij}$ using a logit link. We set
\begin{align}
p_{ij}(a, \lambda) = \begin{cases}\frac{\exp(a + \lambda^2 w_iw_j)}{1 + \exp(a+\lambda^2 w_iw_j)}, & \text{for } i\neq j	\\	0 &\text{ else. }	\end{cases}
\label{eq:connection_probs_other}
\end{align}
The constant $a$ is set to preserve the mean connectivity $p_{\text{mean}}$ through the different heterogeneity levels. This is achieved by (numerically) minimizing the function
\begin{align*} 
f(a) = \left| p_{\text{mean}} - \binom{n}{2}^{-1} \sum_{1\leq i< j\leq n} p_{ij}(a,\sigma^2)\right |.
\end{align*} 
In the end, the resulting probabilities have two values, one if $w_i=w_j$, and another one if $w_i\neq w_j$. Thus, the resulting models are  two-community stochastic block models with two equally sized communities. We set $p_{\text{mean}}=0.3$ and use $\lambda\in\{0,0.5,1,1.5,...,4\}$. Thus, $\lambda=0$ represents the null model. Here, all edge probabilities are $p_{ij}=p_{\text{mean}}$, $i<j=1,...,n$. Increasing values of $\lambda$ yield models with rising heterogeneity: The intra-community connection probabilities increase, whereas the probabilities for inter-community edges decrease. 

\begin{figure}[!t]
\centering
\includegraphics[width=4in, keepaspectratio = TRUE]{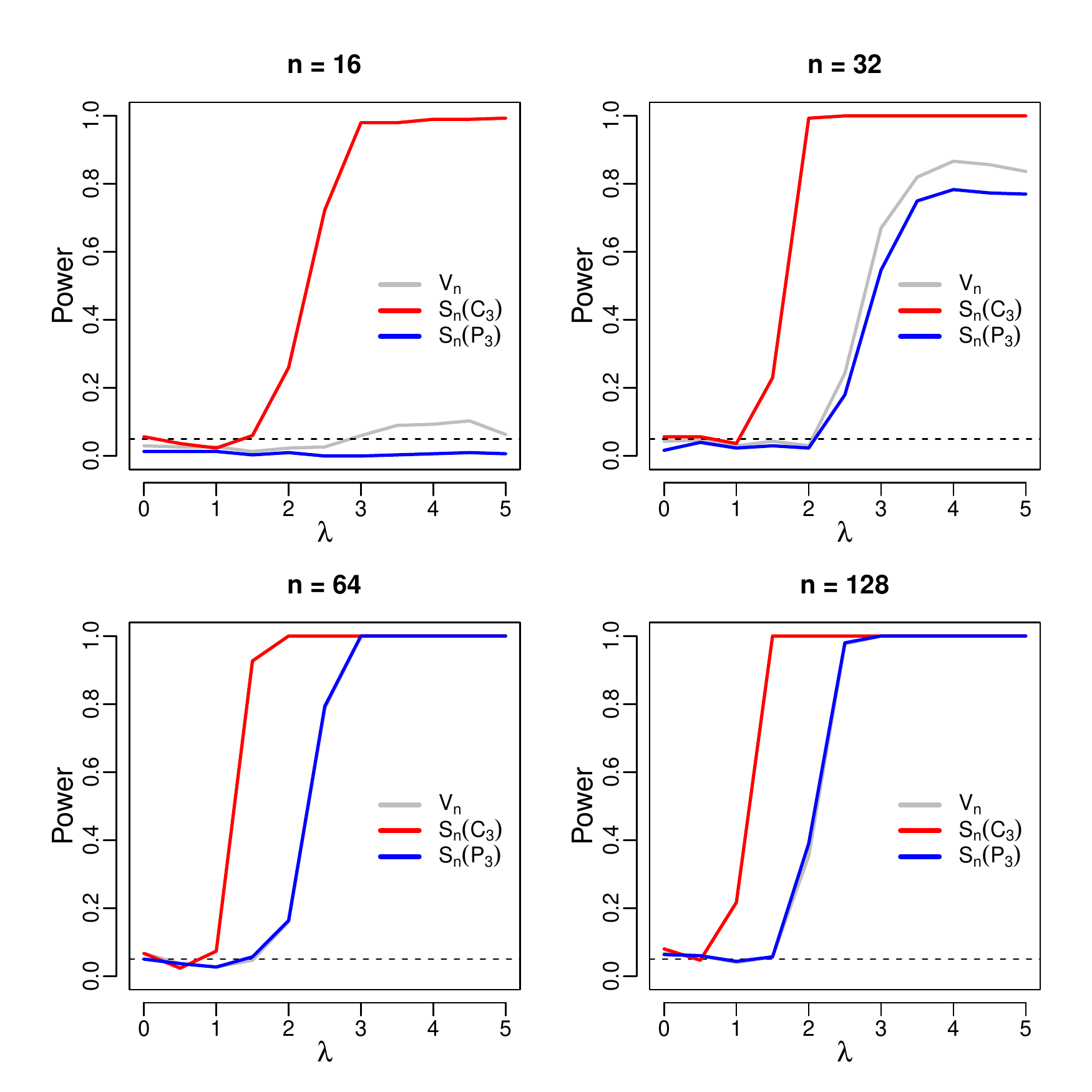}
\caption{Power of the tests $V_n$, $S_n(C_3)$ and $S_n(P_3)$ for the alternative of an SBM with two blocks and $p_{\text{mean}} = \frac{1}{\sqrt{n}}$ for various values for $n$ and $\lambda$.}\label{fig:simSbm2}
\end{figure}

Performance results for $p_{\text{mean}} = \frac{1}{\sqrt{n}}$ are illustrated in Figure \ref{fig:simSbm2}.
Note that these results stand representative for all other values of $p_{\text{mean}}$, since the power of the tests is not directly influenced by the density of the underlying network as explained in Section \ref{sec:power}. Overall, the results are in accordance with our analysis of this mentioned section. It is clearly visible that the tests based on $C_3$ have a superior performance in the underlying SBM cases as it is able to react way more sensitive than both other tests that are mainly based on $P_3$. Whereas $C_3$ is able to detect differences even for low values of the heterogeneity parameter $\lambda$, the two other tests require a rather large deviation from the null model. In this context, the performances get more reliable for all tests with an increasing amount of nodes. For the case of $n = 16$, $P_3$ and $V_n$ are even not at all sensitive for all investigated values of $\lambda$. On a further note, the similar results of these two tests in all situations underline our derivations that $P_3$ is the decisive component of the centered subgraph count representation of $V_n$.  

For our second SBM setup, we construct an SBM with more flexible assumptions to its parameters. Compared to before, it consists of three blocks with random block sizes and varying intra-block probabilities. Formally, we set this up by assigning the weight vector $\bs{w'}$ to the nodes with  
\begin{align*} 
\bs{w'} = (w'_1,...,w'_n) = \bigg( \underbrace{-\frac{1}{2},...,-\frac{1}{2}}_{b_1 \text{ times}}, \underbrace{0,..., 0}_{b_2 \text{ times}}, \underbrace{\frac{1}{4},...,\frac{1}{4}}_{b_3 \text{ times}}\bigg). 
\end{align*}
In this context, $b_k$ with $k = \{1,2,3\}$ denotes the block size of the corresponding block. Each $b_k$ is randomly chosen with the constraint that $b_1 + b_2 + b_3 = n$ and $b_k \geq 2$ for each $k$. The probabilities $p_{ij}$ are then determined analogously to the two-block SBM case explained above.

\begin{figure}[!t]
\centering
\includegraphics[width=4in, keepaspectratio = TRUE]{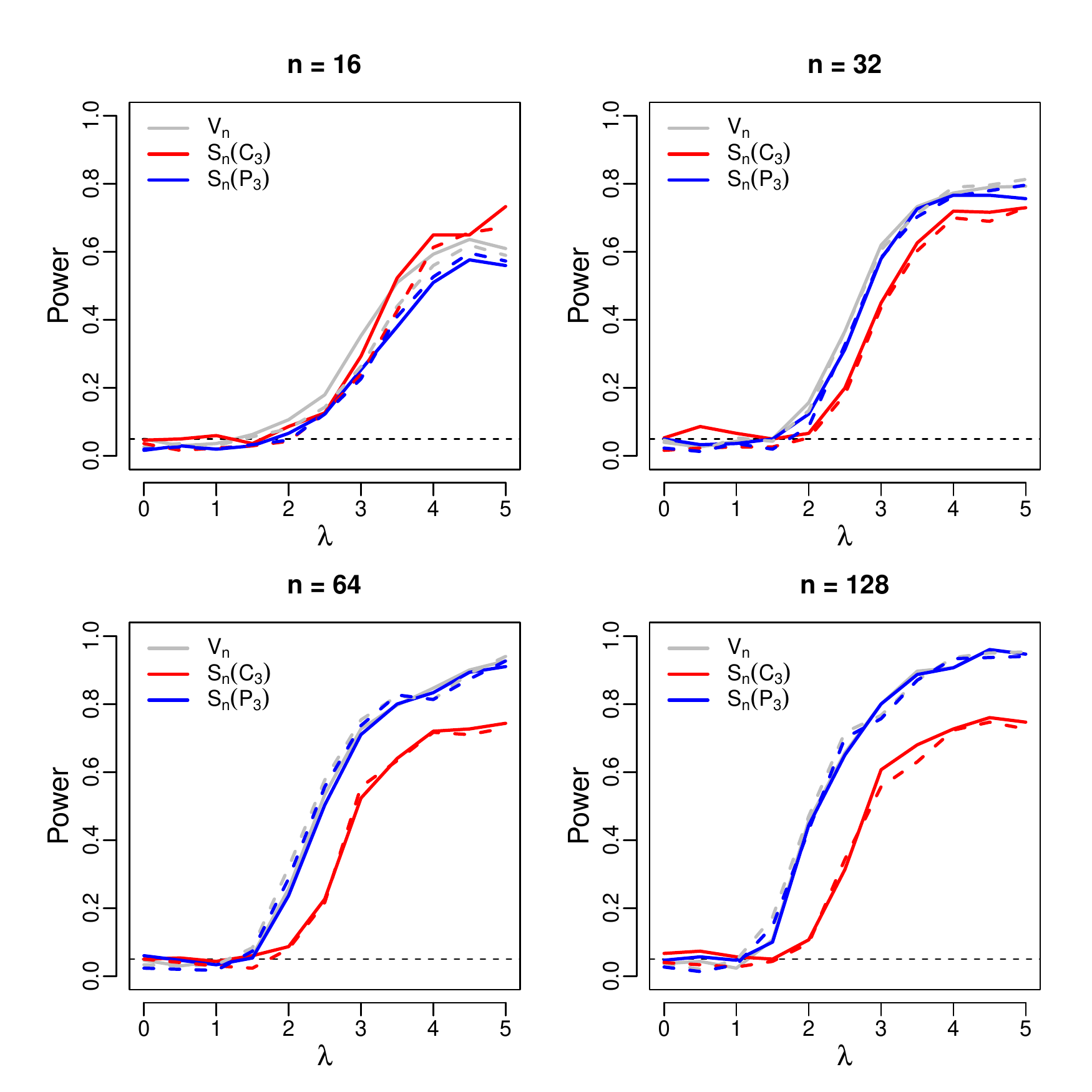}
\caption{Power of the tests $V_n$, $S_n(C_3)$ and $S_n(P_3)$ for the alternative of an SBM with three blocks and $p_{\text{mean}} = \frac{\log{n}}{n}$. The asymptotic versions of the tests are represented by solid lines and the bootstrap version by dashed lines.}\label{fig:sim3SBMsparse}
\end{figure}

The resulting performances are depicted in Figure \ref{fig:sim3SBMsparse} for $p_{\text{mean}} = \frac{\log{n}}{n}$.
Regarding the performance under the null, all tests stick to the desired level $\alpha = 0.05$. Apart from this, the results are, interestingly, quite different to before as both tests based on $P_3$ perform superior compared to the test $C_3$. Apparently, the two-star structure is more sensitive to an increased number of blocks, varying block sizes and varying intra-block probabilities. The bootstrap versions of the tests confirm this behaviour by achieving very similar results to their asymptotic counterparts. The improved performance of $P_3$ and $V_n$ seems quite plausible. As shown in Section \ref{sec:power}, subgraph counts of $P_3$ are, in theory, sensitive to detecting SBMs with the assumptions of Theorem \ref{th:P3}. Although they are sensitive to a lesser extent compared to $C_3$ for the two block case, they might profit from dividing the graph into more blocks as this can be interpreted as a further increase of heterogeneity compared to the ER-model. Furthermore, the varying block sizes and intra-block probabilities might influence this behaviour as well. Interestingly, the test based on $C_3$ performs superior again for denser setups which can be seen in Figure \ref{fig:sim3SBMdense} in Appendix B. In these denser networks, an increased $\lambda$ enables the possibility of larger deviations between the intra-group probabilities and inter-group probabilities which could support the sensitivity of $C_3$.

\subsubsection{Covariate models}
As a further heterogenous alternative, we study covariate models for which the connection probabilities are generated in a way such that vertices with similar properties are more likely to connect than others. This charateristic seems to be a realistic setting for modeling real world networks. Especially in social networks, we expect variables such as the age or the social status to affect whether people know each other or not.
\begin{figure}[!t]
\centering
\includegraphics[width=4in, keepaspectratio = TRUE]{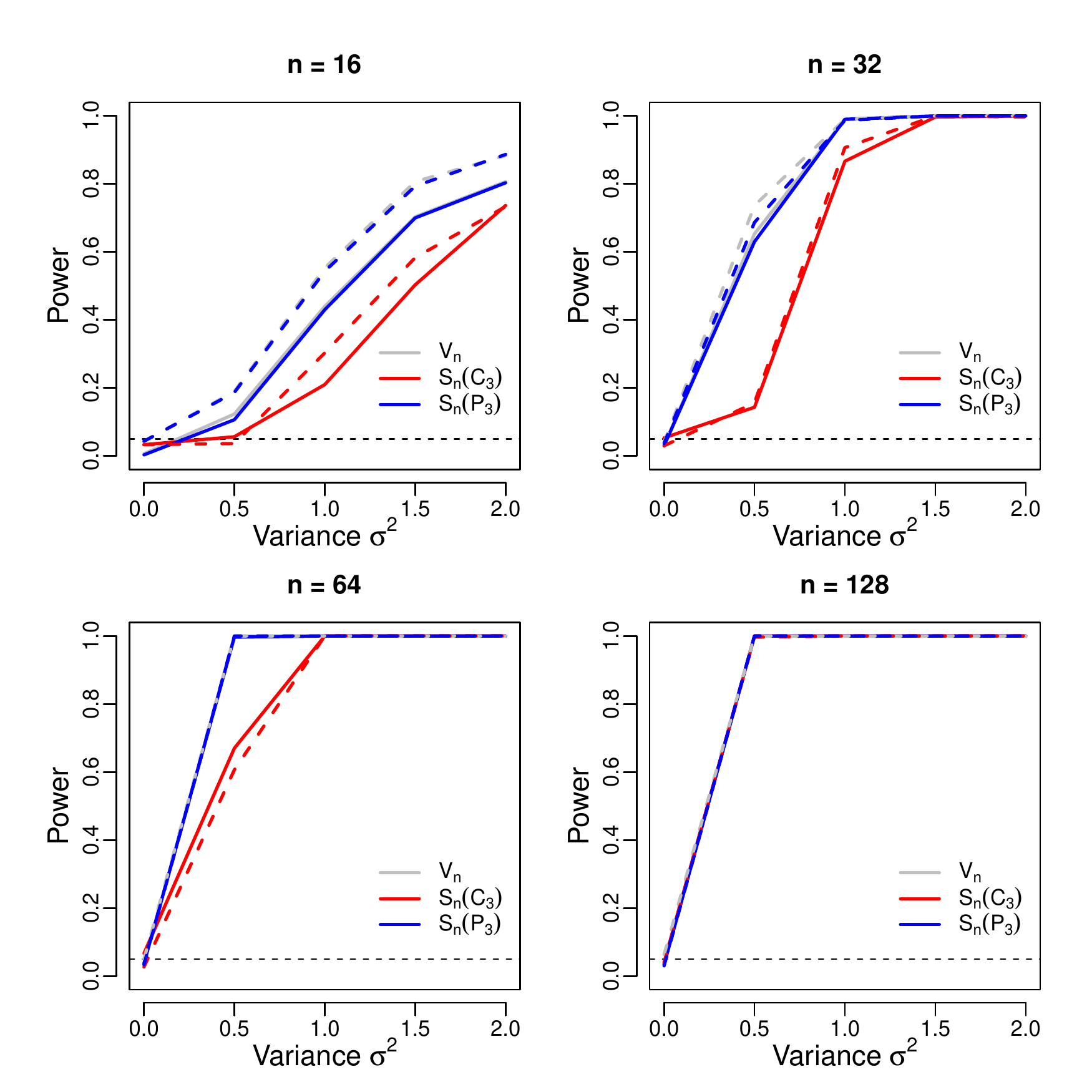}
\caption{Power of the tests $V_n$, $S_n(C_3)$ and $S_n(P_3)$ for the alternative of a covariate model setup with $p_{\text{mean}} = \frac{1}{\sqrt{n}}$. The asymptotic versions of the tests are represented by solid lines and the bootstrap version by dashed lines.}\label{fig:simCovariate}
\end{figure}
To achieve a setting like this, we associate each vertex $v_i\in\{v_1,...,v_n\}$ with a bivariate covariate \begin{align*}
\begin{pmatrix}
x_{i1} \\ x_{i2} 
\end{pmatrix}\sim \mathcal{N}_2\left( \begin{pmatrix}
0 \\0
\end{pmatrix}, \begin{pmatrix}
1 & 0 \\ 0 & 1
\end{pmatrix} \right),
\end{align*} drawn independently from a bivariate standard normal distribution. The covariates are then multiplied by a sequence of variances $\sigma^2\in \{0,0.5,1,1.5,2\}$. This yields versions of the same dataset with different degrees of heterogeneity. The connection probabilities (for given $\sigma^2$ and covariate set) are then calculated according to a logistic model:
\begin{align*}
p_{ij}(a, \sigma^2) = \begin{cases} \frac{\exp(a - \sigma^2|x_{i1} - x_{j1} | - \sigma^2|x_{i2} - x_{j2}|)}{1 + \exp(a - \sigma^2|x_{i1} - x_{j1} | - \sigma^2|x_{i2} - x_{j2}|)}, & \text{ for } i\neq j, \\
 0, & \text{ else. }
\end{cases}
\end{align*}
The constant $a$ is again set to preserve the mean connectivity $p_{\text{mean}}$ and can be determined as before. In this model, $\sigma^2=0$ represents the null hypothesis, each vertex has the same covariate value and the resulting connection probabilities are all equal to $p_{\text{mean}}$. The bigger the variance $\sigma^2$, the more the model deviates from the null model due to the increasing heterogeneity of the covariates. In general, the vertices have a higher probability to connect if their covariates have similar values. Especially, vertices that have unusually large or small covariate values will only be able to draw few connections. The described procedure yields a series of connection probability matrices with increasing heterogeneity.\\
Results are illustrated in Figure \ref{fig:simCovariate} for $p_{\text{mean}} = \frac{1}{\sqrt{n}}$. First of all, the tests reliably stick to the desired level $\alpha=0.05$. Regarding the performance under the alternative, the performance is overall quite good for all tests with advantages for $V_n$ and $P_3$. For larger network sizes, the tests have no problems with detecting the deviation from the null model - even for the smallest investigated value of the heterogeneity parameter $\sigma^2$. Another result is that the bootstrap versions are able to outperform the asymptotic versions for small $n$ and perform equally for larger $n$. This behaviour is also confirmed by the results of most of the other investigated parameter setups that are not reported in the paper. As expected, the derived parametric bootstrap procedure is a promising alternative to the asymptotic version of the proposed class of tests -- particularly for small network sizes.

\section{Conclusion}\label{sec:concl}
Network data is quite complex and its handling is challenging from a statistical point of view. 
An important aspect is the complexity-reduction achieved by using a suitable network model. For a reliable analysis, however, the fitted model should be an adequate representation of the underlying data. In this context, we derived a class of goodness-of-fit tests for networks that serves as a unified approach and contains various formerly proposed tests as special cases. To do so, we used the broad class of graph functionals as test statistics and expand existing theory in order to derive novel asymptotic tests for evaluating if an underlying graph is generated by a homogenous Erdös-Renyi model or some heterogenous alternative. Moreover, we proposed a parametric bootstrap approach that particularly performs favorable in small network size situations compared to the asymptotic tests. We enriched our analysis by the application of our general procedure to three types of test statistics and derived power analysis results for the subgraphs of triangles and two-stars for the popular use case of stochastic block models. We underlined our findings with an extensive simulation study in which we investigated multiple network parameter setups and also studied covariate models as a further heterogenous alternative.

A possible topic for future research is to construct tests for which a heterogenous model $\GnP$ represents the null model in the spirit of \cite{ouadah2020degree}, who exclusively considered the degree variance and without any resampling-based inference. A possible approach could be to try to extend the proof technique for the $\Gnp$ case \citep{janson2011random}, that is based on a continuos time martingale theorem, to the heterogenous case yielding a heterogenous equivalent to the method of higher projections.

\section*{Acknowledgements}
This work was financially supported by the Mercator Research Center Ruhr (MERCUR) with project number PR-2019-0019.

\newpage

\section*{Appendix A: Proofs}

\subsection*{Proof of Theorem \ref{th:boot}}
By assumption, $G\in\mathcal{G}_{HER}$, that is, $G$ is generated by a heterogeneous ER-graph $\mathcal{G}(n,\bs{P})$ with some $\bs{P}=(p_{ij})$, $p_{ij}=p_{ji}$, $p_{ij}\in[0,1]$ for all $i,j$ such that its mean connectivity $p_{\text{mean}}=\binom{n}{2}^{-1} \sum_{1\leq i<j\leq n} p_{n,ij}$ satisfies $p_{\text{mean}}=p_{\text{mean}}(n)\ntoinfty p_0 \in [0,1]$. Then, conditional on $\bs{A} = (A_{ij})_{1\leq i,j\leq n}$, $G^*$ is generated by a homogeneous ER-graph $\mathcal{G}(n,\widehat p)$, where $\widehat{p} =\binom{n}{2}^{-1} \sum_{1\leq i<j\leq n} A_{ij}$. Furthermore, we have that, under the null, $X_n = \sum_{H\in \mathcal{H}} a_n(H) S_n(H)$, which is dominated by some family of connected graphs $\mathcal{H}_0$ such that for all $H\in\mathcal{H}_0$, it holds $np_{\text{mean}}^{r(H)} \ntoinfty \infty$. Hence, under both the null and the alternative, we immediately get $X_n^* = \sum_{H\in \mathcal{H}_0} a_n^*(H) S_n^*(H)$, which is dominated by the same family of connected graphs $\mathcal{H}_0$, and where $a_n^*(H)=a_n(H;\widehat p)$ for $a_n(H)=a_n(H;p)$.

Hence, in view of Theorem \ref{thm:as_normality}, we have to check, whether for all $H\in\mathcal{H}_0$, we have that $n\widehat p^{r(H)} \ntoinfty \infty$, the coefficients
\begin{align}\label{boot_condition1}
\widehat b(H) = \sup_n \frac{n^{n(H)/2} \widehat p^{m(H)/2} a_n^*(H) }{\sqrt{\Var^*(X_n^*)}}
\end{align}
are finite and satisfy
\begin{align}\label{boot_condition2}
\sum_{H\in \mathcal{H}_0} \widehat b^2(H) |\text{aut}(H)| < \infty.
\end{align}
in probability, respectively. For this purpose, let us consider $\widehat{p} =\binom{n}{2}^{-1} \sum_{1\leq i<j\leq n} A_{ij}$ in more detail. To be more precise, as $p_{ij}$, and consequently also the distribution of $A_{ij}$, are allowed to depend on $n$, we will write $A_{n,ij}$ and $p_{n,ij}$ in the following. For the expectation, we get
\begin{align}\label{phat_ex}
\E(\widehat p)=\E\left(\binom{n}{2}^{-1} \sum\limits_{1\leq i<j\leq n} A_{n,ij}\right) = \binom{n}{2}^{-1} \sum\limits_{1\leq i<j\leq n} \E(A_{n,ij}) = \binom{n}{2}^{-1} \sum\limits_{1\leq i<j\leq n} p_{n,ij}=p_{\text{mean}}.
\end{align}
For the variance, as the edges are formed independently according to Bernoulli distributions with connection probabilities $p_{n,ij}$, we have
\begin{align}\label{phat_var}
\Var(\widehat p) = \binom{n}{2}^{-2}\sum\limits_{1\leq i<j\leq n} \Var(A_{n,ij}) = \binom{n}{2}^{-2}\sum\limits_{1\leq i<j\leq n} p_{n,ij}(1-p_{n,ij}). 
\end{align}
Hence, to prove that $n\widehat p^{r(H)} \ntoinfty \infty$ holds in probability, using $1\leq r(h)<\infty$, we get
\begin{align*}
n\widehat p^{r(H)} = n\left(\widehat p-p_{mean}+p_{mean}\right)^{r(H)} = np_{mean}^{r(H)}+n\sum_{j=1}^{r(H)}\binom{r(H)}{j}(\widehat p-p_{mean})^j p_{mean}^{r(H)-j}
\end{align*}
As $np^{r(H)} \ntoinfty \infty$ by assumption, it remains to argue that all other terms on the last right-hand side are bounded in probability. This follows easily from \eqref{phat_var} leading to 
\begin{align}\label{phat_rate}
\widehat p-p_{mean} = O_P\left(\sqrt{\binom{n}{2}^{-2}\sum\limits_{1\leq i<j\leq n} p_{n,ij}(1-p_{n,ij})}\right) \leq O_P\left(\frac{\min\{p_{mean},1-p_{mean}\}}{n}\right)\leq O_P\left(\frac{1}{n}\right)
\end{align}
due to $\binom{n}{2}=n^2$ and $1-p_{n,ij}\leq 1$ and $p_{n,ij}\leq 1$. As $p_{mean}\in[0,1]$, we have also $p_{mean}^{r(H)-j}\in[0,1]$ for all $j=1,\ldots,r(H)$ such that
\begin{align*}
n\sum_{j=1}^{r(H)}\binom{r(H)}{j}(\widehat p-p_{mean})^j p_{mean}^{r(H)-j}=O_P(1).
\end{align*}
To show that \eqref{boot_condition1} is bounded in probability, note that $X_n^* = \sum_{H\in \mathcal{H}_0} a_n^*(H) S_n^*(H)$ with $a_n^*(H)=a_n(H;\widehat p)$ and 
\begin{align*}
\Var^*(X_n^*) = \sum_{H\in\mathcal{H}_0} a_n^{*2}(H) \Var(S_n^*(H)),
\end{align*}
where, according to Proposition \ref{lemma:traits_of_sg_counts},
\begin{align*}
\Var^*(S_n^*(H)) = |\text{aut}(H)|(n)_{n(H)} (\widehat p(1-\widehat p))^{m(H)}.
\end{align*}
By plugging-in, we get
\begin{align*}
\widehat b(H) = \sup_n \frac{n^{n(H)/2} \widehat p^{m(H)/2} a_n(H) }{\sqrt{\Var^*(X_n^*)}} = \sup_n \frac{n^{n(H)/2} \widehat p^{m(H)/2} a_n(H) }{\sqrt{|\text{aut}(H)|(n)_{n(H)} (\widehat p(1-\widehat p))^{m(H)}}}
\end{align*}
which remains bounded in probability by similar arguments as above due to \eqref{boot_assumption1} and making use of \eqref{phat_rate}. Finally, using the same arguments, \eqref{boot_condition2} follows from \eqref{boot_assumption2}.	\hfill $\square$


\subsection*{Proof of Corollary \ref{cor:estimated_moments_Vn}}
(i) Making use of \eqref{phat_ex} and \eqref{phat_var} under the null of a homogeneous ER graph, from \eqref{phat_rate}, we get the rate
\begin{align}\label{phat_order}
\widehat p-p = O_P\left(\frac{p}{n}\right).
\end{align}
Consequently, we also have
\begin{align*}
\hat{p}(1-\hat{p}) - p(1-p) =& \hat{p}(1-\hat{p}) - \hat{p}(1-p) + \hat{p}(1-p) - p(1-p) = \hat{p}(p-\hat p) + (\hat{p}-p)(1-p)	\\
=& (\hat{p}-p+p)(p-\hat p) + (\hat{p}-p)(1-p) = p(p-\hat p)-(\hat p-p)^2+(\hat{p}-p)(1-p)	\\
=& \mathcal{O}_{\mathbb{P}}\left(\frac{p^2}{n}\right) + \mathcal{O}_{\mathbb{P}}\left(\frac{p^2}{n^2}\right) + \mathcal{O}_{\mathbb{P}}\left(\frac{p}{n}\right) = \mathcal{O}_{\mathbb{P}}\left(\frac{p}{n}\right)
\end{align*}
We then consider the following expansion:
	\begin{align*}
	\frac{V_n - \E_{\hat{p}}(V_n)}{\sqrt{\Var_{\hat{p}}(V_n)}} & =  \frac{\sqrt{\Var_p(V_n)}}{\sqrt{\Var_{\hat{p}}(V_n)}} \left(\frac{V_n - \E_{p}(V_n)}{\sqrt{\Var_{p}(V_n)}} - \frac{\E_{\ph}(V_n)-\E_{p}(V_n)}{\sqrt{\Var_p(V_n)}}\right) . 
	\end{align*} 
The difference of the expectations in the numerator of the second term in brackets is bounded in probability due to
	\begin{align*}
	\E_{\ph}(V_n)-\E_{p}(V_n) \,= \,&  \frac{(n-1)(n-2)}{n} \Big( \ph(1-\ph) - p(1-p) \Big) = \landau(n) \cdot \mathcal{O}_{\mathbb{P}}\left(\frac{p}{n}\right) = \mathcal{O}_{\mathbb{P}}(p).
	\end{align*}
As the variance $\Var_p(V_n)$ has order $np^2$, and $np\rightarrow \infty$ by assumption, we obtain
	\begin{align*}
	\frac{\E_{\ph}(V_n)-\E_{p}(V_n)}{\sqrt{\Var_p(V_n)}} \convP 0, ~\text{ as well as }~ \frac{\sqrt{\Var_p(V_n)}}{\sqrt{\Var_{\hat{p}}(V_n)}} \convP 1,
	\end{align*}
	by similar arguments using $\widehat p-p = O_P(\frac{p}{n})$. Combining these two convergence results and applying Slutzky's lemma, we get that $\frac{V_n - \E_{\hat{p}}(V_n)}{\sqrt{\Var_{\hat{p}}(V_n)}}$ has the same asymptotic distribution as $\frac{V_n - \E_{p}(V_n)}{\sqrt{\Var_{p}(V_n)}}$ stated in \eqref{Vn_asymp}.
(ii) By the same arguments as above and used to obtain \eqref{Vn1}, \eqref{Vn2}, \eqref{Vn2a}, \eqref{Vn3}, \eqref{Vn4}, \eqref{Vn5} and \eqref{Vn6}, all conditions of Theorem \ref{th:boot} can be shown. Exemplarily, we have 
\begin{align*}
\widehat b(H_2) = \sqrt{\frac{2}{(1-\hat{p})^2}},
\end{align*}
which is finite in probability, because $\hat p\rightarrow p_0\in[0,1)$ by assumption such that $(1-\hat{p})^2$ is bounded away from zero with probability tending to one. In particular, we have
\begin{align*}
\prob(\hat{p} = 1) = & \prob(A_{ij} = 1 ~ \forall \, 1\leq i<j \leq n) = p^{\binom{n}{2}} \xrightarrow[n\to\infty]{} 0
\end{align*}
for all $p=p_n\rightarrow p_0\in[0,1)$.	\hfill	$\square$


\subsection*{Proof of Theorem \ref{rem:estmoments}}
For readability purposes let $|\text{aut}_n(H)| := |\text{aut}(H)|(n)_{n(H)}$ From Proposition \ref{prop:normality_of_subgraph_counts} and due to the representation 
\begin{align*}
\frac{ \hat{S}_n(H)}{ \sqrt{|\text{aut}_n(H)|(\hat{p}(1-\hat{p}))^{m(H)} } } = \sqrt{\frac{|\text{aut}_n(H)|(p(1-p))^{m(H)}}{|\text{aut}_n(H)|(\hat{p}(1-\hat{p}))^{m(H)}}}\left( \frac{S_n(H)}{\sqrt{|\text{aut}_n(H)|(p(1-p))^{m(H)}}} + \frac{\hat{S}_n(H) - S_n(H)}{\sqrt{|\text{aut}_n(H)|(p(1-p))^{m(H)}}} \right),
\end{align*}
to prove the claimed result, it remains to show
\begin{align*}
\textnormal{(I)} \quad \frac{\hat{S}_n(H) - S_n(H)}{\sqrt{|\text{aut}_n(H)| (p(1-p))^{m(H)} } } \convP 0, \quad \text{and} \quad  \textnormal{(II)} \quad  \frac{({p}(1-{p}))^{m(H)}}{(\hat{p}(1-\hat{p}))^{m(H)}} \convP 1 .
\end{align*}
For condition $\text{(I)}$, note that $\hat{S}_n(H)$ is obtained from $S_n(H)$ by replacing $\mathbb{P}(e\in E(G))$ in \eqref{Sn_def1} by $\widehat p$, which is a consistent estimator of $p=\mathbb{P}(e\in E(G))$ under the null of an ER graph. Hence, for deriving the asymptotic properties of $\hat{S}_n(H)-S_n(H)$ under the null, it is natural to re-write $\hat{S}_n(H)$ in terms of centered subgraph counts. We have
\begin{align*}
\widehat{S}_n(H) =& \sum_{\widetilde{H}\in \text{iso}_n(H)} \prod_{e\in E(\widetilde{H})} (\mathbbm{1}\{e\in  E(G)\}-\widehat{p}) = \sum_{\widetilde{H}\in \text{iso}_n(H)} \prod_{e\in E(\widetilde{H})} (\mathbbm{1}\{e\in  E(G)\}-p+p-\widehat{p})	\\
=& \sum_{\widetilde{H}\in \text{iso}_n(H)} \prod_{e\in E(\widetilde{H})} \left\{(\mathbbm{1}\{e\in  E(G)\}-p)+(p-\widehat{p})\right\}	\\
=& \sum_{\widetilde{H}\in \text{iso}_n(H)} \sum_{\widetilde J\subseteq \widetilde H: E(\widetilde J)\in\mathcal{P}(E(\widetilde H))}
\left\{\left(\prod_{e\in E(J)}(\mathbbm{1}\{e\in  E(G)\}-p)\right)(p-\widehat{p})^{m(\widetilde H)-m(\widetilde J)}\right\}	\\
=& \sum_{J\subseteq H: E(J)\in\mathcal{P}(E(H))}\left(\sum_{\widetilde{J}\in \text{iso}_n(J)} \prod_{e\in E(\widetilde{J})} (\mathbbm{1}\{e\in  E(G)\}-p)\right) (p-\widehat p)^{m(H)-m(J)}	\\
=& \sum_{J\subseteq H: E(J)\in\mathcal{P}(E(H))}S_n(J) (p-\widehat p)^{m(H)-m(J)}	\\
=& S_n(H) + \sum_{J\subsetneq H: E(J)\in\mathcal{P}(E(H))}S_n(J) (p-\widehat p)^{m(H)-m(J)}
\end{align*}
where $\mathcal{P}(E(\widetilde H))$ denotes the power set of $E(\widetilde H)$, that is, the set of all subsets of $E(\widetilde H)$. Hence, altogether, we get
\begin{align}
\frac{\hat{S}_n(H) - S_n(H)}{\sqrt{|\text{aut}_n(H)| (p(1-p))^{m(H)}}} = \frac{\sum_{J\subsetneq H: E(J)\in\mathcal{P}(E(H))}S_n(J) (p-\widehat p)^{m(H)-m(J)}}{\sqrt{|\text{aut}_n(H)| (p(1-p))^{m(H)}}}.
\end{align}
Then, making use of \eqref{Sn_CLT} for all $J\subsetneq H$ and \eqref{phat_order}, the last right-hand side becomes a term of order
\begin{align}
O_P\left(\frac{\sum_{J\subsetneq H: E(J)\in\mathcal{P}(E(H))}\left(n^{n(J)/2}p^{m(J)/2}\right) \left((\frac{p}{n})^{m(H)-m(J)}\right)}{\sqrt{|\text{aut}_n(H)| (p(1-p))^{m(H)}}}.\right)=o_P(1)
\end{align}
as $\sum_{J\subsetneq H: E(J)\in\mathcal{P}(E(H))}$ is a finite sum and $m(H)>m(J)$.		
Condition {\normalfont(II)} follows directly from the consistency of $\hat{p}$.	\hfill	$\square$

\subsection*{Proof of Lemma \ref{lemma:phat}}
First, Equation \eqref{eq:phat_subgraph} is obtained due to
\begin{align*}
	\widehat{p} = \binom{n}{2}^{-1} \sum_{1\leq i < j\leq n}A_{ij} = \binom{n}{2}^{-1} \sum_{1\leq i < j\leq n}(A_{ij} - p) + p = \binom{n}{2}^{-1} S_n(P_2) + p,
\end{align*}
where $P_2$ is a graph with two nodes and one edge. In order to obtain the dominating family for $\hat{p}^3$, we have
\begin{align} \label{eq:phat_sn_3}
	\widehat{p}^3  = \left(\binom{n}{2}^{-1} S_n(P_2) + p\right)^3  = \left(\binom{n}{2}^{-1} S_n(P_2)\right)^3 + 3\left(\binom{n}{2}^{-1} S_n(P_2)\right)^2p + 3\binom{n}{2}^{-1} S_n(P_2)p^2 + p^3 
\end{align}
As can be seen, $\widehat{p}^3$ decomposes into four summands, where the first two terms
\begin{align*}
	(A) := \left(\binom{n}{2}^{-1} S_n(P_2)\right)^3	\quad \text{ and}	\quad	(B) := 3p\left(\binom{n}{2}^{-1} S_n(P_2)\right)^2
\end{align*}
are of lower order than the third term $(C) := 3\binom{n}{2}^{-1} S_n(P_2)p^2$ due to the following. From Proposition \ref{prop:normality_of_subgraph_counts}, we have
\begin{align*}
	\left(p(1-p)\right)^{-1/2}n^{-1}S_n(P_2) \convD N(0, 1)
\end{align*}
and thus, we can directly conclude that $\left(p(1-p)\right)^{-1/2}n^{-1}S_n(P_2)=\mathcal{O}_{\prob}(1)$ holds such that $S_n(P_2)=\mathcal{O}_{\prob}(\left(p(1-p)\right)^{1/2}n)$. Thus, we have $\binom{n}{2}^{-1}\binom{n}{2}^{-1}S_n(P_2)=\mathcal{O}_{\prob}(\left(p(1-p)\right)^{1/2}n^{-1})$, which yields $(A) = \mathcal{O}_{\prob}(\left(p(1-p)\right)^{3/2}n^{-3})$ and $(B) = \mathcal{O}_{\prob}(p^2(1-p)n^{-2})$ as well as $(C) = \mathcal{O}_{\prob}(p^{5/2}(1-p)^{1/2}n^{-1})$. Hence, $(C)$ becomes the leading term of $\widehat{p}^3$ in \eqref{eq:phat_sn_3}. Similarly, it is possible to show that the variances of $(A)$ and $(B)$ are of lower order than the variance of $(C)$ such that 
we have 
\begin{align*}
	\frac{\Var(\widehat{p}^3))}{	\Var\left(3\binom{n}{2}^{-1}p^2 S_n(P_2)\right) } \xrightarrow[n\to\infty]{} 1
\end{align*}
and the dominating family for $\widehat{p}^3$ is indeed formed by $\mathcal{H}_0 = \{P_2\}$, which proves Lemma \ref{lemma:phat}. \hfill	$\square$

\subsection*{Proof of Corollary \ref{cor:TnC3}}

We are interested in the asymptotic distribution of the graph functional
\begin{align} \label{eq:Tn_centered_proof}
	T_n(C_3) - \binom{n}{3}\hat{p}^3.
\end{align}
Equation \eqref{eq:Tn_centered} follows by plugging in the decomposition \eqref{eq:phat_sn_3} into \eqref{eq:TnC3_decomposition}. Calculation of its variance then directly yields \eqref{eq:var_stand}.
Thus, the dominating family for the graph functional \eqref{eq:Tn_centered_proof} is \begin{align*}
	\mathcal{H}_0 = \{C_3, P_3\}, \text{ with coefficients } a_n(C_3) = 1 \text{ and } a_n(P_3) = p
\end{align*}
The corresponding coefficients $b(H)$ are given by
\begin{align*}
	b(C_3) =  \sup_n \frac{n^{3/2}p^{3/2}}{\sqrt{\Var(T_n(C_3) - \binom{n}{3}\hat{p}^3)}}.
\end{align*}		
As the leading term of the variance in the denominator is $\frac{1}{6}\left(1 - \frac{3}{n} + \frac{2}{n^2}\right)((1-p)^3 + 3p(1-p)^2)$, the finiteness of the supremum above is implied by
\begin{align*}
& \lim_{n\to\infty} \frac{1}{\sqrt{\frac{1}{6}\left(1 - \frac{3}{n} + \frac{2}{n^2}\right)((1-p)^3 + 3p(1-p)^2) + \frac{const.}{n} }} \\
= & \frac{1}{\sqrt{\frac{1}{6}((1-p_0)^3 + 3p_0(1-p_0)^2)}},
\end{align*}
which is finite for $p_0\in(0,1)$. Analogously, we get
\begin{align*}
	b(P_3) = & \sup_n \frac{n^{3/2}p}{\sqrt{\Var(T_n(C_3) - \binom{n}{3}\hat{p}^3)}} =   \frac{1}{\sqrt{\frac{1}{6}(p_0(1-p_0)^3 + 3p_0^2(1-p_0)^2)}}
\end{align*}
Thus,
\begin{align*}
	\sum_{H\in\mathcal{H}_0} b(H)^2|\text{aut}(H)| = b(C_3)^2 + 3b(P_3) < \infty
\end{align*}
and Theorem \ref{thm:as_normality} is applicable. Corollary \ref{cor:TnC3} directly follows from this.	\hfill	$\square$


\subsection*{Proof of Theorem \ref{th:C3}}
Following is the proof for statement (a).\\
It is $\E_{\text{ER}}(T_n(C_3)) = \binom{n}{3} p^3$ and $\E_{\text{SBM}}(T_n(C_3)) = \left[2 \binom{\nicefrac{n}{2}}{3} p^3_{\text{intra}} + \left[\left(\binom{n}{3} - 2 \binom{\nicefrac{n}{2}}{3}\right)p_{\text{intra}}p_{\text{inter}}^2\right]\right] $. \\
Hence, it yields
\begin{align*}
& \E_{\text{ER}}(T_n(C_3)) - \E_{\text{SBM}}(T_n(C_3)) = \binom{n}{3} p^3 - \left[2 \binom{\nicefrac{n}{2}}{3} p^3_{\text{intra}} + \left[\left(\binom{n}{3} - 2 \binom{\nicefrac{n}{2}}{3}\right)p_{\text{intra}}p_{\text{inter}}^2\right]\right] \\
& = \binom{n}{3} p^3 - \binom{n}{3} p_{\text{intra}}p_{\text{inter}}^2 - 2\binom{\nicefrac{n}{2}}{3} p^3_{\text{intra}} + 2\binom{\nicefrac{n}{2}}{3}p_{\text{intra}}p_{\text{inter}}^2\\
&= \binom{n}{3}(p^3 -  p_{\text{intra}}p_{\text{inter}}^2) + 2\binom{\nicefrac{n}{2}}{3}(p_{\text{intra}}p_{\text{inter}}^2 - p_{\text{intra}}^3) \\
&= \binom{n}{3} \left(\left(\frac{1}{2}\frac{n-2}{n-1}(p_{\text{intra}}+p_{\text{inter}})\right)^3 - p_{\text{intra}}p_{\text{inter}}^2\right) + 2\binom{\nicefrac{n}{2}}{3}(p_{\text{intra}}p_{\text{inter}}^2 - p_{\text{intra}}^3)\\
&= \binom{n}{3} \frac{1}{8} \left(\frac{n-2}{n-1}\right)^3(p_{\text{intra}}^3 + 3p_{\text{intra}}^2p_{\text{inter}} + 3p_{\text{intra}}p_{\text{inter}}^2+ p_{\text{inter}}^3) - \binom{n}{3} p_{\text{intra}}p_{\text{inter}}^2 
+ 2\binom{\nicefrac{n}{2}}{3}(p_{\text{intra}}p_{\text{inter}}^2 - p_{\text{intra}}^3)\\
&= \binom{n}{3} \frac{1}{8} \left(\frac{n-2}{n-1}\right)^3(p_{\text{intra}}^3 + 3p_{\text{intra}}^2p_{\text{inter}} + p_{\text{inter}}^3) + \binom{n}{3}\frac{1}{8} \frac{n-2}{n-1} 3p_{\text{intra}}p_{\text{inter}}^2 - \binom{n}{3} p_{\text{intra}}p_{\text{inter}}^2 
+ 2\binom{\nicefrac{n}{2}}{3}(p_{\text{intra}}p_{\text{inter}}^2 - p_{\text{intra}}^3)
\end{align*}
The last right-hand side is asymptotically equivalent to
\begin{align*}
& \binom{n}{3} \frac{1}{8} (p_{\text{intra}}^3 + 3p_{\text{intra}}^2p_{\text{inter}} + p_{\text{inter}}^3) - \frac{5}{8} p_{\text{intra}}p_{\text{inter}}^2 
+ 2\binom{\nicefrac{n}{2}}{3}(p_{\text{intra}}p_{\text{inter}}^2 - p_{\text{intra}}^3)\\
&= \frac{1}{8} \binom{n}{3} (p_{\text{intra}}^3 + 3p_{\text{intra}}^2p_{\text{inter}} + p_{\text{inter}}^3 - 5p_{\text{intra}}p_{\text{inter}}^2) + 2\binom{\nicefrac{n}{2}}{3}(p_{\text{intra}}p_{\text{inter}}^2 - p_{\text{intra}}^3)\\
&= \frac{1}{8} \binom{n}{3} p_{\text{intra}}^3 + 
   \frac{3}{8} \binom{n}{3} p_{\text{intra}}^2p_{\text{inter}} + 
   \frac{1}{8} \binom{n}{3} p_{\text{inter}}^3 - 
   \frac{5}{8} \binom{n}{3} p_{\text{intra}}p_{\text{inter}}^2 + 
   2\binom{\nicefrac{n}{2}}{3}p_{\text{intra}}p_{\text{inter}}^2 - 
   2\binom{\nicefrac{n}{2}}{3} p_{\text{intra}}^3 \\
&= \left(\frac{1}{8}\binom{n}{3}-2\binom{\nicefrac{n}{2}}{3}\right)p_{\text{intra}}^3
   + \frac{3}{8} \binom{n}{3} p_{\text{intra}}^2p_{\text{inter}}
   + \frac{1}{8} \binom{n}{3} p_{\text{inter}}^3
   + \left(2\binom{\nicefrac{n}{2}}{3}-\frac{5}{8}\binom{n}{3}\right)p_{\text{intra}}p_{\text{inter}}^2,
\end{align*}	
which, using \eqref{hilfsresultat}, is asymptotically equivalent to
\begin{align*}
&\xrightarrow{} \left(\binom{\nicefrac{n}{2}}{3} - 2\binom{\nicefrac{n}{2}}{3} \right)p_{\text{intra}}^3 + 
\binom{\nicefrac{n}{2}}{3} p_{\text{intra}}^2p_{\text{inter}} +
\binom{\nicefrac{n}{2}}{3} p_{\text{inter}}^3 + 
\left(2\binom{\nicefrac{n}{2}}{3} - 5\binom{\nicefrac{n}{2}}{3} \right) p_{\text{intra}}p_{\text{inter}}^2 \\
&= -\binom{\nicefrac{n}{2}}{3}p_{\text{intra}}^3 +  
\binom{\nicefrac{n}{2}}{3}p_{\text{inter}}^3 +
3\binom{\nicefrac{n}{2}}{3} p_{\text{intra}}^2p_{\text{inter}} - 
3\binom{\nicefrac{n}{2}}{3} p_{\text{intra}}p_{\text{inter}}^2 \\
&= \binom{\nicefrac{n}{2}}{3} (p_{\text{inter}} - p_{\text{intra}})^3  < 0
\end{align*}
with $p_{\text{intra}} \geq p_{\text{inter}}$ and
\begin{align}\label{hilfsresultat}
\frac{\binom{n}{3}}{\binom{\nicefrac{n}{2}}{3}} &= \frac{n!}{\nicefrac{n}{2}!} \frac{(\nicefrac{n}{2} - 3)!}{(n - 3)!} = \frac{n(n - 1)(n - 2)}{\nicefrac{n}{2} (\nicefrac{n}{2} - 1) (\nicefrac{n}{2} - 1) (\nicefrac{n}{2} - 2)} = \frac{(n - 1)(n - 2)}{(\nicefrac{n}{4} - \nicefrac{1}{2})(\nicefrac{n}{2} - 2)} \\
&= \frac{(n - 1)(n - 2)}{\nicefrac{1}{4}(n - 2)\nicefrac{1}{2}(n - 4)} = \frac{n - 1}{\nicefrac{1}{8}(n - 4)} = 8\frac{n - 1}{n - 4} \ntoinfty 8
\end{align}
Under the assumed conditions, the amount of expected triangles in a SBM is larger than in a homogenous ER-model. The difference in expectation increases for larger values of $p_\text{intra}$.


Now let us move to the proof of statement (b). The end result was already shown in the main text and the derivations for the homogenous ER-model case as well. Still missing are the derivations for the SBM case. In this context, we have:
For $I_n(C_3)$ and $I_n(E_3)$, we have analogously to the proof of statement (a)
\begin{align*}
\E_\text{SBM}(I_n(C_3)) &= 2 \binHalf \pra^3 + \left[\left(\bin - 2\binHalf \right) \pra \per^2\right]\\
\E_\text{SBM}(I_n(E_3)) &= 2 \binHalf (1 - \pra^3) + \left[\left(\bin - 2\binHalf \right) (1 -\pra) (1 - \per)^2\right].
\end{align*}
For $I_n(P_3)$ and $I_n(D_3)$ it becomes more difficult as the present edges can exist between the blocks and/or in the blocks. This results in the possibilities
\begin{align*}
I_n(P_3)_{\text{SBM}} &= \begin{cases}
3 \binHalf (1 - \pra)\pra^2, & \text{all corresponding nodes in the same block} \\
(*), & \text{corresponding nodes from two different blocks.}
\end{cases}\\
\text{with} (*) &= \begin{cases}
2 \binom{\nicefrac{n}{2}}{2}\frac{n}{2} \pra \per (1 - \per), & \text{missing link between nodes of different blocks} \\
\binom{\nicefrac{n}{2}}{2}\frac{n}{2} \per^2 (1 - \pra), & \text{missing link between the nodes of the same block.}
\end{cases}\\
I_n(D_3)_{\text{SBM}} &= \begin{cases}
3 \binHalf (1 - \pra)^2 \pra, & \text{all corresponding nodes in the same block} \\
(**), & \text{corresponding nodes from two different blocks.}
\end{cases}\\
\text{with} (**) &= \begin{cases}
2 \binom{\nicefrac{n}{2}}{2}\frac{n}{2} (1 - \pra) \per (1 - \per), & \text{existing link between nodes of different blocks} \\
\binom{\nicefrac{n}{2}}{2}\frac{n}{2} \pra (1 - \per)^2, & \text{existing link between nodes of the same block,}
\end{cases}
\end{align*} 
yielding
\begin{align*}
\E_\text{SBM}(I_n(P_3)) &= 6 \binHalf \pra^2 (1 - \pra) + 2n \binom{\nicefrac{n}{2}}{2} \pra \per (1 - \per) + n \binom{\nicefrac{n}{2}}{2} \per^2 (1 - \pra).  \\
\E_\text{SBM}(I_n(D_3)) &= 6 \binHalf \pra (1 - \pra)^2 + 2n \binom{\nicefrac{n}{2}}{2} (1 - \pra) \per (1 - \per) + n \binom{\nicefrac{n}{2}}{2} \pra (1 - \per)^2.  
\end{align*}
Hence, we have
\begin{align*}
\E_\text{SBM}(S_n(C_3)) &= (1 - p_{\text{mean}})^3 \left[2 \binHalf \pra^3 + \left[\left(\bin - 2\binHalf \right) \pra \per^2\right] \right] \\
& + (1 - p_{\text{mean}})^2 (-p_{\text{mean}}) \left[6 \binHalf \pra^2 (1 - \pra) + 2n \binom{\nicefrac{n}{2}}{2} \pra \per (1 - \per) + n \binom{\nicefrac{n}{2}}{2} \per^2 (1 - \pra) \right] \\
& + (1 - p_{\text{mean}})p_{\text{mean}}^2 \left[6 \binHalf \pra (1 - \pra)^2 + 2n \binom{\nicefrac{n}{2}}{2} (1 - \pra) \per (1 - \per) + n \binom{\nicefrac{n}{2}}{2} \pra (1 - \per)^2 \right] \\
& + (-p_{\text{mean}})^3 \left[2 \binHalf (1 - \pra^3) + \left[\left(\bin - 2\binHalf \right) (1 -\pra) (1 - \per)^2\right]\right].
\end{align*}
\hfill	$\square$


\subsection*{Proof of Theorem \ref{th:P3}}
Let us start with the proof of statement (b). The procedure is similar to the proof of statement (b) of Theorem \ref{th:C3} that is presented in the main text. We can again use the decomposition of $E(S_n(H))$ into the linear combination of the expected values of induced subgraphs for all possible subgraph shapes consisting of the same node amount as $H$. Compared to before, only the coefficients change resulting in
\begin{align*}
\E(S_n(P_3)) =& \E(I_n(C_3)  3(1 - p)^2) + \E(I_n(P_3)  [(1-p)^2 + 2(1-p)(-p)])+ \E(I_n(D_3)  [(-p)^2 + 2(1-p)(-p)]) \\ &+  \E(I_n(E_3) 3(-p)^2)\\
=& 3(1 - p)^2\E(I_n(C_3)) + [(1-p)^2 + 2(1-p)(-p)]\E(I_n(P_3))\\&+ [(-p)^2 + 2(1-p)(-p)]\E(I_n(D_3)) +  3(-p)^2\E(I_n(E_3)).
\end{align*}
It is again $\E_{\text{ER}}(S_n(P_3) = 0$, since we considered centered (!) subgraph counts. Plugging in the derivations from the proof of Theorem \ref{th:C3} in the formula above, it yields for the SBM case
\begin{align*}
&\E_\text{SBM}(S_n(P_3)) = 3(1 - p_{\text{mean}})^2 \left[2 \binHalf \pra^3 + \left[\left(\bin - 2\binHalf \right) \pra \per^2\right] \right] \\
& + \left[(1 - p_{\text{mean}})^2 + 2(-p_{\text{mean}})(1-p_{\text{mean}})\right] \left[6 \binHalf \pra^2 (1 - \pra) + 2n \binom{\nicefrac{n}{2}}{2} \pra \per (1 - \per) + n \binom{\nicefrac{n}{2}}{2} \per^2 (1 - \pra) \right] \\
& + \left[(-p_{\text{mean}})^2 + 2(-p_{\text{mean}})(1-p_{\text{mean}})\right]  \left[6 \binHalf \pra (1 - \pra)^2 + 2n \binom{\nicefrac{n}{2}}{2} (1 - \pra) \per (1 - \per) + n \binom{\nicefrac{n}{2}}{2} \pra (1 - \per)^2 \right] \\
& + 3(-p_{\text{mean}})^2 \left[2 \binHalf (1 - \pra^3) + \left[\left(\bin - 2\binHalf \right) (1 -\pra) (1 - \per)^2\right]\right].
\end{align*}

Now, let us move on to the proof of statement (a). As for the proof of Theorem \ref{th:C3}, we investigate the difference of the expected values of $T_n(P_3)$ for the ER-model and the SBM model with the mentioned assumptions. It is
\begin{align*}
& \E_{\text{ER}}(T_n(P_3)) - \E_{\text{SBM}}(T_n(P_3))  \\
&= 3 \bin (1-p) p^2 - \left[6 \binHalf \pra^2 (1-\pra) + 2n \binHalb \pra \per (1-\per) +n \binHalb \per^2 (1-\pra)\right] \\
&= 3 \bin (1-p) p^2 - 6 \binHalf \pra^2 (1-\pra) - 2n \binHalb \pra \per (1-\per) - n \binHalb \per^2 (1-\pra).
\end{align*}
Let us first have a look at the first term that comes from the ER-model. We reformulate $p$ in dependence of $\pra$ and $\per$ with $p = \pra = \per$ which results in $p = \frac{1}{2}(\pra + \per)$. Note that this connection is also asymptotically (!) valid for the SBM case. It therefore is
\begin{align*}
& \E_{\text{ER}}(T_n(P_3)) - \E_{\text{SBM}}(T_n(P_3)) = \\
&= 3\bin \left[ \left(1 - \frac{1}{2}\pra - \frac{1}{2}\per\right)\frac{1}{4}(\pra^2 + 2\pra\per + \per^2) \right]\\
&- 6 \binHalf \pra^2 (1-\pra) - 2n \binHalb \pra \per (1-\per) - n \binHalb \per^2 (1-\pra)\\
&= \frac{3}{4} \bin \left[\pra^2 + 2\pra\per + \per^2- \frac{1}{2} \pra^3 - \pra^2 \per - \frac{1}{2}\pra\per^2 - \frac{1}{2} \pra^2\per - \pra\per^2 - \frac{1}{2} \per^3 \right] \\
&- 6 \binHalf \pra^2 (1-\pra) - 2n \binHalb \pra \per (1-\per) - n \binHalb \per^2 (1-\pra) \\
\end{align*}
The last right-hand side can be written as
\begin{align*}
&= 6 \binHalf \frac{n-1}{n-4} \left[\pra^2 + 2\pra\per + \per^2 - \frac{1}{2} \pra^3 - \frac{1}{2} \per^3 - \frac{3}{2} \pra^2\per - \frac{3}{2}\pra\per^2 \right] \\
&- 6 \binHalf \pra^2 (1-\pra) - 2n \binHalb \pra \per (1-\per) - n \binHalb \per^2 (1-\pra) \\[5pt]
&= 6 \frac{n-1}{n-4} \binHalf \left(-\frac{1}{2} \per^3 - \frac{3}{2} \pra^2\per\right) + 6 \frac{n-1}{n-4} \binHalf \pra^2 - 6 \binHalf \pra^2 + 6 \binHalf \pra^3 \\
&- 3 \frac{n-1}{n-4} \binHalf \pra^3 + \pra\per \left(12 \frac{n-1}{n-4} \binHalf - 2n \binHalb \right) \\
&+ \pra\per^2 \left(-9 \frac{n-1}{n-4} \binHalf + 3n \binHalb \right) + \per^2 \left(6 \frac{n-1}{n-4} \binHalf - n \binHalb\right) \\
&= 6 \frac{n-1}{n-4} \binHalf \left(-\frac{1}{2} \per^3 - \frac{3}{2} \pra^2\per\right) + 6 \frac{n-1}{n-4} \binHalf \pra^2 - 6 \binHalf \pra^2 + 6 \binHalf \pra^3 \\
&- 3 \frac{n-1}{n-4} \binHalf \pra^3 + \pra\per \left(12 \frac{n-1}{n-4} \frac{\nicefrac{n}{2} - 2}{3} \binHalb - 2n \binHalb \right) \\
&+ \pra\per^2 \left(-9 \frac{n-1}{n-4} \frac{\nicefrac{n}{2} - 2}{3} \binHalb + 3n \binHalb \right) + \per^2 \left(6 \frac{n-1}{n-4} \frac{\nicefrac{n}{2} - 2}{3} \binHalb - n \binHalb\right)
\end{align*}
This results in
\begin{align*}
&\pra^2 \cdot 6 \binHalb \frac{\nicefrac{n}{2} - 2}{3} \left(\frac{n-1}{n-4} - 1\right) + \pra^3 \cdot 6 \binHalb \frac{\nicefrac{n}{2} - 2}{3} \left(1- \frac{1}{2} \frac{n-1}{n-4}\right) \\ 
&- \pra^2\per \cdot 9 \binHalb \frac{\nicefrac{n}{2} - 2}{3} \frac{n-1}{n-4} + \pra\per \cdot \left(\frac{n-1}{n-4} (2n - 8) \binHalb - 2n \binHalb\right) \\ 
&- \per^3 \cdot 3 \frac{\nicefrac{n}{2} - 2}{3} \binHalb \frac{n-1}{n-4}  + \pra\per^2 \left(- \frac{n-1}{n-4} \left(\frac{3n}{2} - 6\right) \binHalb + 3n \binHalb\right) + \per^2 \left(\frac{n-1}{n-4}(n-4) \binHalb - n \binHalb\right)\\
&= \binHalb \left[\pra^2 \cdot 6 \frac{\nicefrac{n}{2} - 2}{3} \left(\frac{n-1}{n-4} - 1\right)\right. + \pra^3 \cdot 6 \frac{\nicefrac{n}{2} - 2}{3} \left(1- \frac{1}{2} \frac{n-1}{n-4}\right) \\ 
&- \pra^2\per \cdot 9 \frac{\nicefrac{n}{2} - 2}{3} \frac{n-1}{n-4} + \pra\per \cdot \left(\frac{n-1}{n-4} (2n - 8) - 2n \right) \\ 
&- \per^3 \cdot 3 \frac{\nicefrac{n}{2} - 2}{3} \frac{n-1}{n-4}  + \pra\per^2 \left(- \frac{n-1}{n-4} \left(\frac{3n}{2} - 6\right) + 3n \right) + \left.\per^2 \left(\frac{n-1}{n-4}(n-4) - n \right)\right] \\
&= \binHalb \left[\pra^2 \cdot 2 (\frac{n}{2} - 2) \left(\frac{3}{n-4}\right)\right. + \pra^3 \cdot 2 (\frac{n}{2} - 2) \left( \frac{n-7}{2(n-4)}\right) \\ 
&- \pra^2\per \cdot 3 (\frac{n}{2} - 2) \frac{n-1}{n-4} + \pra\per \cdot \left(2(n-1) - 2n \right) \\ 
&- \per^3 \cdot \left(\frac{n}{2} - 2\right) \frac{n-1}{n-4} + \pra\per^2 \cdot \left(- \frac{3(n-1)}{2} + 3n \right) + \left.\per^2 \cdot (-1) \right] \\
&= \binHalb \left[\pra^2 \cdot 3 + \pra^3 \cdot \frac{n-7}{2}+ \pra^2\per \cdot \frac{3-3n}{2} + \pra\per \cdot (-2) \right.\\
&+ \left.\per^3 \cdot \left(\frac{1-n}{2}\right) + \pra\per^2 \cdot \left(\frac{3n+3}{2}\right) + \per^2 \cdot (-1) \right]
\end{align*}
For interpretation purposes, we set $\pra = x \cdot \per$ w.l.o.g. resulting in
\begin{align*}
&= \binHalb \left[(x \cdot \per)^2 \cdot 3 + (x \cdot \per)^3 \cdot \frac{n-7}{2}+ (x \cdot \per)^2\per \cdot \frac{3-3n}{2} + (x \cdot \per)\per \cdot (-2) \right.\\
&+ \left.\per^3 \cdot \left(\frac{1-n}{2}\right) + (x \cdot \per)\per^2 \cdot \left(\frac{3n+3}{2}\right) + \per^2 \cdot (-1) \right] \\
&= \binHalb \left[x^2 \cdot \per^2 \cdot 3 + x^3 \cdot \per^3 \cdot \frac{n-7}{2}+ x^2 \cdot \per^3 \cdot \frac{3-3n}{2} + x \cdot \per^2 \cdot (-2) \right.\\
&+ \left.\per^3 \cdot \left(\frac{1-n}{2}\right) + x \cdot \per^3 \cdot \left(\frac{3n+3}{2}\right) + \per^2 \cdot (-1) \right]\\[5pt]
&= \binHalb \left[\per^3 (x^3 \cdot \frac{n-7}{2} + x^2 \cdot \frac{3-3n}{2} + x \cdot \frac{3n+3}{2} + \frac{1-n}{2}) + \per^2 (3x^2 - 2x - 1) \right]\\
&= \binHalb \left[\per^3\left(\frac{n}{2}(x^3-3x^2+3x-1)-\frac{1}{2}(x-1)(7x^2+4x+1)\right) + \per^2(x-1)(3x+1)\right] \\[5pt]
&= \binHalb \left[\frac{1}{2} \per^3 ( n (x-1)^3 - (x-1)(7x^2+4x+1)) + \per^2(x-1)(3x+1)\right]\\
&= \binHalb (x-1) \left[\frac{1}{2} \per^3 (n(x-1)^2 - (7x^2+4x+1)) + \per^2(3x+1)\right].
\end{align*}
Our general assumption in SBMs is $\pra > \per$ which in this case is equivalent to $x>1$.
Obviously, the coefficient $\binHalb (x-1)$ is larger than 0 in this case as is the last term in the bracket, i.e. $\per^2(3x+1)$. Remaining is the term $\frac{1}{2} \per^3 (n(x-1)^2 - (7x^2+4x+1))$. This term can be negative for a few setups with small $n$. However, for most realistic network setups and particularly for $n \to \infty$ this term is also larger than 0 because $(n-7)x^2$ increases larger than $(-2n - 4)x$ decreases. Thus, we have
\begin{align*}
\E_{\text{ER}}(T_n(P_3)) - \E_{\text{SBM}}(T_n(P_3)) = \underbrace{\binHalb (x-1)}_{>0} \left[\frac{1}{2}\per^3 (\underbrace{n(x-1)^2 - (7x^2+4x+1))}_{>0 \text{\space for \space} n \to \infty} + \underbrace{\per^2(3x+1)}_{>0}\right]
\end{align*}
\hfill	$\square$

\newpage

\section*{Appendix B: Additional tables and figures}

\begin{figure}[h]
\centering
\includegraphics[width=4.25in, keepaspectratio = TRUE]{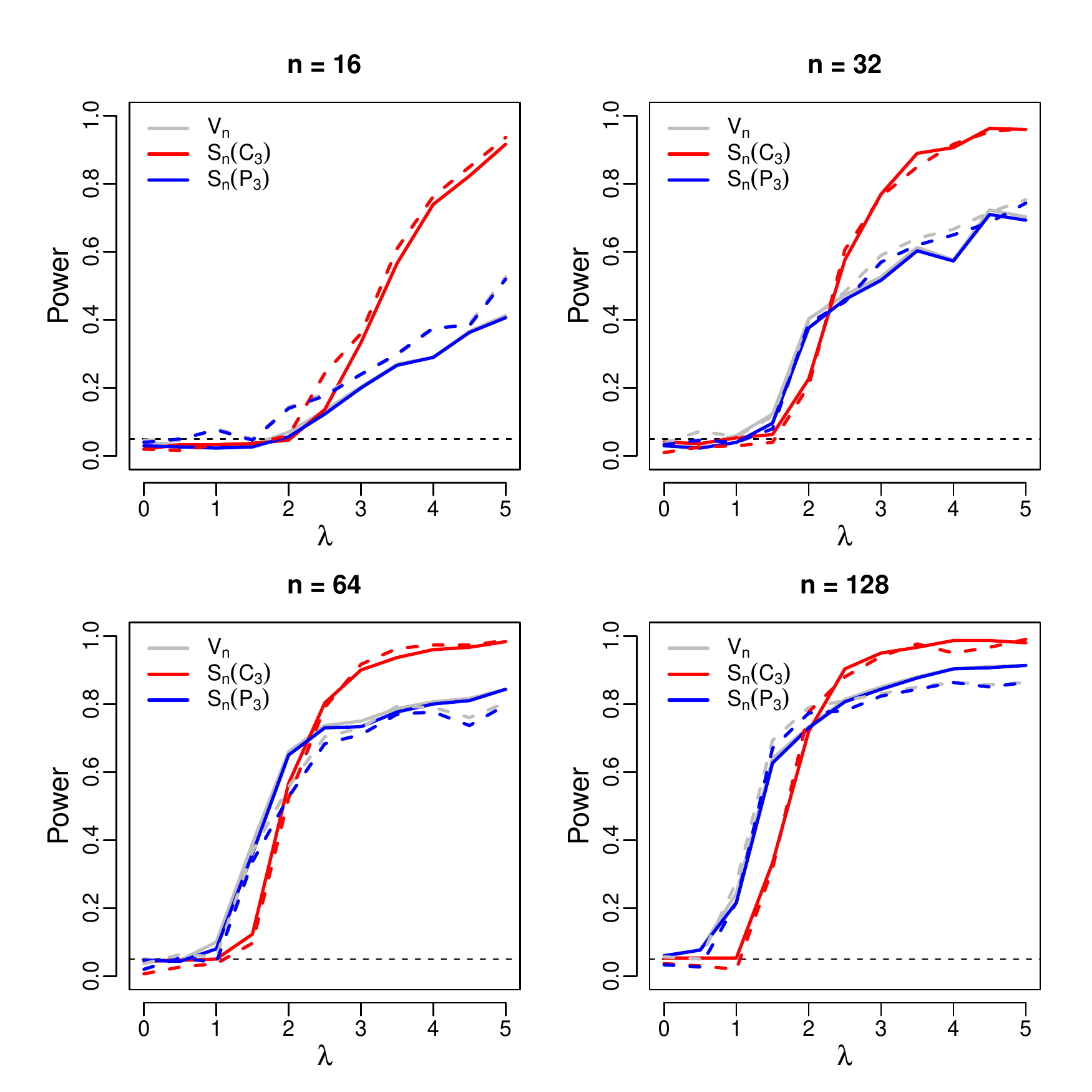}
\caption{Power of the tests $V_n$, $S_n(C_3)$ and $S_n(P_3)$ for the alternative of a SBM setup with three blocks with $p_{\text{mean}} = \frac{\log{n}}{\sqrt{n}}$. The asymptotic versions of the tests are represented by solid lines and the bootstrap version by dashed lines.}\label{fig:sim3SBMdense}
\end{figure}



\end{document}